\documentclass[fleqn,usenatbib]{mnras}

\usepackage[T1]{fontenc}

\DeclareRobustCommand{\VAN}[3]{#2}
\let\VANthebibliography\thebibliography
\def\thebibliography{\DeclareRobustCommand{\VAN}[3]{##3}\VANthebibliography}

\usepackage{graphicx}	
\usepackage{amsmath}	
\usepackage{amssymb}	
\usepackage{multirow}   
\usepackage{newtxtext,newtxmath}
\usepackage{gensymb}

\graphicspath{{figs/}}

\setcitestyle{notesep={}}

\newcommand{\fluxunit}{erg\,s$^{-1}$\,cm$^{-2}$}

\title[H$\alpha$ detection in Dusty Submm Galaxy HDF850.1]{Mapping dusty galaxy growth at $z>5$ with FRESCO: Detection of H$\alpha$ in submm galaxy HDF850.1 and the surrounding overdense structures}
\author[Herard-Demanche et al.]{Thomas Herard-Demanche$^{1}$\thanks{email: herard@strw.leidenuniv.nl},
Rychard J. Bouwens$^{1}$, Pascal A. Oesch$^{2,3}$, Rohan P. Naidu$^{4}$, Roberto Decarli$^{5}$,
\newauthor
Erica J. Nelson$^{6}$, Gabriel Brammer$^{2}$, Andrea Weibel$^{3}$, Mengyuan Xiao$^{3}$, Mauro Stefanon$^{7,8}$, Fabian Walter$^{9}$, 
\newauthor
Jorryt Matthee$^{10}$, Romain A. Meyer$^{3,9}$, Stijn Wuyts$^{11}$, Naveen Reddy$^{12}$, Pablo Arrabal Haro$^{13}$, 
\newauthor
Helmut Dannerbauer$^{14,15}$, Alice E. Shapley$^{16}$, John Chisholm$^{17}$, Pieter van Dokkum$^{18}$, Ivo Labbe$^{19}$,
\newauthor
Garth Illingworth$^{20}$, Daniel Schaerer$^{3}$, Irene Shivaei$^{21}$ \\ 
$^{1}$Leiden Observatory, Leiden University, NL-2300 RA Leiden, Netherlands\\
$^{2}$Cosmic Dawn Center (DAWN), Niels Bohr Institute, University of Copenhagen, Jagtvej 128, K\o benhavn N, DK-2200, Denmark\\
$^{3}$Department of Astronomy, University of Geneva, Chemin Pegasi 51, 1290 Versoix, Switzerland\\
$^{4}$MIT Kavli Institute for Astrophysics and Space Research, 77 Massachusetts Ave., Cambridge, MA 02139, USA\\
$^{5}$INAF - Osservatorio di Astrofisica e Scienza dello Spazio di Bologna, Via Gobetti 93/3, 40129, Bologna, Italy\\
$^{6}$Department for Astrophysical and Planetary Science, University of Colorado, Boulder, CO 80309, USA\\
$^{7}$Departament d'Astronomia i Astrof\`isica, Universitat de Val\`encia, C. Dr. Moliner 50, E-46100 Burjassot, Val\`encia,  Spain\\
$^{8}$Unidad Asociada CSIC "Grupo de Astrof\'isica Extragal\'actica y Cosmolog\'ia" (Instituto de F\'isica de Cantabria - Universitat de Val\`encia)\\
$^{9}$Max-Planck-Institut für Astronomie, Königstuhl 17, D-69117 Heidelberg, Germany\\
$^{10}$Department of Physics, ETH Z{\"u}rich, Wolfgang-Pauli-Strasse 27, Z{\"u}rich, 8093, Switzerland\\
$^{11}$Department of Physics, University of Bath, Claverton Down, Bath, BA2 7AY, UK\\
$^{12}$Department of Physics and Astronomy, University of California, Riverside, 900 University Avenue, Riverside, CA 92521, USA\\
$^{13}$NSF's National Optical-Infrared Astronomy Research Laboratory, 950 N. Cherry Avenue, Tucson, AZ 85719, USA\\
$^{14}$Instituto Astrofísica de Canarias (IAC), E-38205 La Laguna, Tenerife, Spain\\
$^{15}$Universidad de La Laguna, Dpto. Astrofísica, E-38206 La Laguna, Tenerife, Spain\\
$^{16}$Department of Physics \& Astronomy, University of California, Los Angeles, 430 Portola Plaza, Los Angeles, CA 90095, USA\\
$^{17}$Department of Astronomy, The University of Texas at Austin, 2515 Speedway, Stop C1400, Austin, TX 78712-1205, USA\\
$^{18}$Astronomy Department, Yale University, 52 Hillhouse Ave, New Haven, CT 06511, USA\\
$^{19}$Centre for Astrophysics and Supercomputing, Swinburne University of Technology, Melbourne, VIC 3122, Australia\\
$^{20}$Department of Astronomy and Astrophysics, University of California, Santa Cruz, CA 95064, USA\\
$^{21}$Centro de Astrobiolog\'{i}a (CAB), CSIC-INTA, Ctra. de Ajalvir km 4, Torrej\'{o}n de Ardoz, E-28850, Madrid, Spain\\
}

\date{Accepted XXX. Received YYY; in original form 2023 September 8}

\pubyear{2023}

\begin{document}
\label{firstpage}
\pagerange{\pageref{firstpage}--\pageref{lastpage}}
\maketitle
\begin{abstract}
We report the detection of a 13$\sigma$ H$\alpha$ emission line from HDF850.1 at $z=5.188\pm0.001$ using the FRESCO NIRCam F444W grism observations.  Detection of H$\alpha$ in HDF850.1 is noteworthy, given its high far-IR luminosity, substantial dust obscuration, and the historical challenges in deriving its redshift.  HDF850.1 shows a clear detection in the F444W imaging data, distributed  between a northern and southern component, mirroring that seen in [CII] from the Plateau de Bure Interferometer.  Modeling the SED of each component separately, we find that the northern component has a higher mass, star formation rate (SFR), and dust extinction than the 
southern component.  The observed H$\alpha$ emission appears to arise entirely from the less-obscured southern component and shows a similar $\Delta$v$\sim$+130 km/s velocity offset to that seen for [CII] relative to the source systemic redshift.  Leveraging H$\alpha$-derived redshifts from FRESCO observations, we find that HDF850.1 is forming in one of the richest environments identified to date at $z>5$, with 100 $z=5.17$--5.20 galaxies distributed across 10 structures and a $\sim$(15 cMpc)$^3$ volume.  Based on the evolution of analogous structures in cosmological simulations, the $z=5.17$--5.20 structures seem likely to collapse into a single $>$10$^{14}$ $M_{\odot}$ cluster by $z\sim0$.  Comparing galaxy properties forming within this overdensity with those outside, we find the masses, SFRs, and $UV$ luminosities inside the overdensity to be clearly higher.  The prominence of H$\alpha$ line emission from HDF850.1 and other known highly-obscured $z>5$ galaxies illustrates the potential of NIRCam-grism programs to map both the early build-up of IR-luminous galaxies and overdense structures.
\end{abstract}



\section{Introduction} \label{sec:intro}

One of the most important frontiers in extragalactic astronomy has been understanding the formation and evolution of massive galaxies in the early universe.  Despite the many significant insights that have been gained into both the build-up of the star formation rates (SFRs) and stellar masses of $UV$-bright galaxies from both space and ground-based telescopes \citep[e.g.,][]{MadauDickinson14,Davidzon17,Stefanon21,Bouwens21,Harikane22}, it is essential we achieve an equally complete census of star formation from obscured galaxies 
given how significantly obscured star formation contributes to galaxies with masses $>$10$^{10}$ $M_{\odot}$ \citep[e.g.][]{Reddy06,Reddy08,Whitaker2017}.  

In addition to the great strides made by ALMA in this area \citep[e.g.,][]{Casey2021_MORA,Smail2021,Bouwens2022_REBELS,Dayal2022}, a JWST is further revolutionizing this science area thanks to its
sensitive near-IR photometric and spectroscopic capabilities to
5$\mu$m and beyond, sampling bright Balmer and Paschen series line like H$\alpha$, Pa$\alpha$, and Pa$\beta$ to $z\sim7$ \citep[e.g.][]{Helton2023,AM2023,Reddy2023_Paschen}.  Through the identification of bright line emission from e.g. H$\alpha$ (e.g., the candidate $z=5.58$ dusty galaxy shown in \citealt{Oesch2023_FRESCO}), one can assemble substantial spectroscopic samples of far-IR luminous, dusty star-forming galaxies in the early universe with JWST.  Particularly useful for this endeavor are NIRCam grism observations that allow for a probe of H$\alpha$ emission over the full morphology of sources over a $\sim$9 arcmin$^2$ NIRCam field, facilitating redshift determinations even when the escape of H$\alpha$ occurs over just an isolated region.

In this paper, we report on the successful detection of H$\alpha$ emission from HDF850.1, one of the first sub--millimeter galaxies to be identified in the high-redshift universe, leveraging new grism observations from the First Reionization Era Spectroscopically Complete Observations (FRESCO: \citealt{Oesch2023_FRESCO}) program.  HDF850.1 was initially identified
using sensitive sub--millimeter observations over the Hubble Deep Field (HDF) North with the SCUBA camera \citep{Hughes1998}, but no optical counterpart could be identified in the deepest HST images available at the time.  Based on the available information, HDF\,850.1 appeared to have a redshift z\,$>$\,2, implying that the star formation rate of this single source could rival the total star formation activity of all coeval galaxies in the HDF combined.  This discovery triggered a $>$10-year effort to improve the location of the source from the coarse position provided by the 17" FWHM SCUBA beam, to search for possible counterparts, and to pin down its redshift \citep{Richards1999,Downes1999,Dunlop2004,Wagg2007,Cowie2009}.  

A redshift for HDF\,850.1 was finally obtained through a millimeter line--scan obtained at the IRAM Plateau de Bure Interferometer (PdBI) by \citet{Walter2012} who detected multiple lines in the rest--frame sub--millimeter ([CII] as well as multiple CO lines), thus unambigously pinpointing the redshift to z\,=\,5.183. This study also revealed that the system is part of a fairly significant galaxy overdensity at these redshifts.  Higher resolution imaging of the [CII] line emission from HDF\,850.1 \citep{Neri2014} provided evidence that HDF\,850.1 is a merging system. 



\begin{figure*}
	\includegraphics[width=\linewidth]{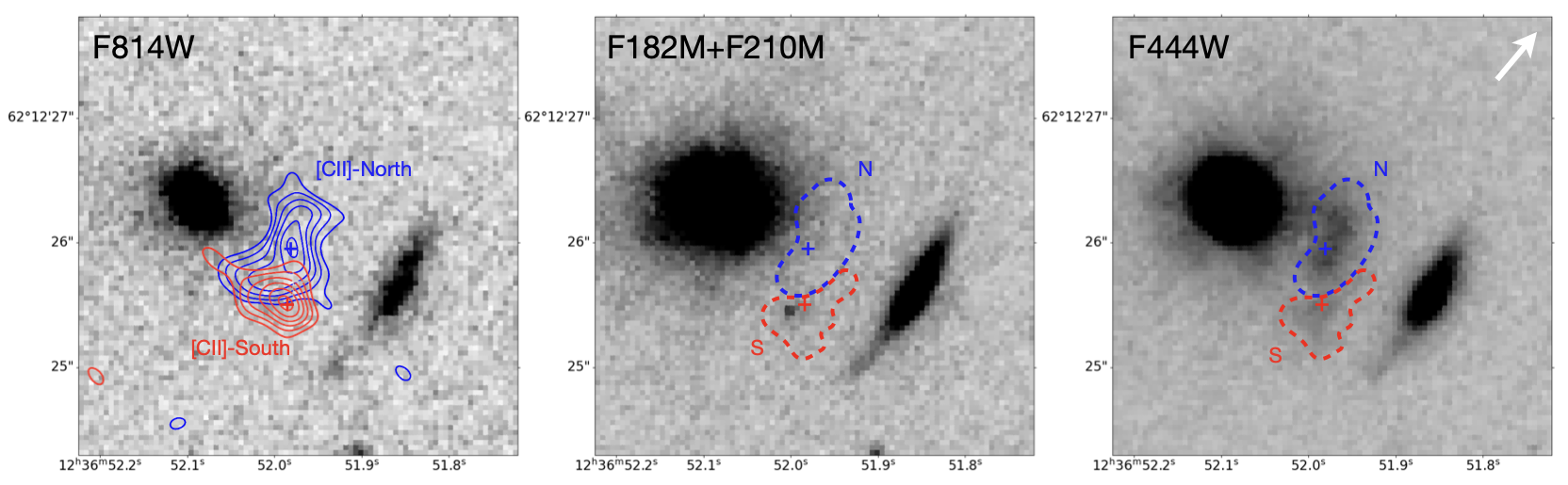}
    \caption{Images centered on HDF850.1 from both \textit{HST} in the F814W band (\textit{left panel}) and \textit{JWST}/NIRCam in the F182M+F210M (\textit{center panel}) and F444W bands (\textit{right panel}).  The blue and red solid lines contours in the left panel show the 3, 4, 5, 6, 7, and 8$\sigma$ contours for [CII] emission from the northern and southern components, respectively, of HDF850.1 as derived by \citet{Neri2014} using PdBI.  Crosses give the spatial position where the [CII] emission for each component reaches a peak.  The dashed contours shown in the center and right panels delineate the regions used in our extractions of near-IR spectra for the northern (\textit{blue}) and southern (\textit{red}) components.  These two regions have been designed to mirror the shape and morphology of the two components in the F444W imaging data. The white arrow in the upper-right corner of the right panel shows the dispersion direction of the NIRCam grism. \label{fig:stamps}}
\end{figure*}

In addition to enabling the detection of the H$\alpha$ line from HDF850.1, the new NIRCam imaging observations from FRESCO allow us to measure the continuum flux from HDF850.1 in the rest-optical while facilitating a modeling of the overall SED of both components 
of the source.  Moreover, the NIRCam grism observations allow for the identification of a large number of H$\alpha$ emitters over a $\sim$62 arcmin$^2$ FRESCO area in the 
GOODS North field, allowing us to map out the $z=5.17$--5.20 overdensity much more extensively than had been possible in earlier work by \citet{Walter2012}, \citet{AH2018}, and \citet{Calvi2021}.

We organize the paper as follows.  In \S\ref{sec:data}, we provide a summary of the observational data we utilize for this analysis and describe our procedures for constructing multi-wavelength catalogs.  In \S\ref{sec:results}, we describe our discovery of a detection of H$\alpha$ from HDF850.1, the photometry we derive for the two components of HDF850.1 and inferences based on SED modeling, 
and then present information on both the structure of the $z=5.17$--5.20
overdensity in which HUDF850.1 resides, and on the characteristics of
the large number of other member galaxies.  In \S\ref{sec:potential}, we briefly discuss the potential of NIRCam grism surveys for mapping out the build-up of massive galaxies in the early universe given the discovery of H$\alpha$ emission from HDF850.1 and a significant fraction of other well-known far-IR luminous $z>5$ galaxies over the GOODS-North field.  In \S\ref{sec:summary}, we include a short summary of our primary results and provide a brief look to the future.  For consistency with previous work, we express all quantities using the so-called concordance cosmology with $\Omega_m = 0.3$ and $\Omega_{\Lambda} = 0.7$, $H_0 = 70$ km/s/Mpc.  Stellar masses and SFRs are presented assuming a \citet{Chabrier03} IMF.  All magnitudes are presented using the AB magnitude system \citep{Oke83}.

\section{Data}
\label{sec:data}

\subsection{FRESCO NIRCam Grism and Imaging Data}
\label{sec:datadesc}

In this analysis we make use of NIRCam data obtained by the JWST FRESCO survey (GO-1895; see \citealt{Oesch2023_FRESCO} for details). FRESCO covered both the GOODS-South and the GOODS-North fields with $\sim$62 arcmin$^2$ of NIRCam/grism spectroscopy in the F444W filter. This coverage is achieved with two 2$\times$4 mosaics of NIRCam/grism observations with significant column overlap in order to maximize the wavelength coverage over the field. Only the GRISMR was used due to overhead costs. The maximal wavelength coverage is from 3.8 to 5.0 $\mu$m, which is achieved over 73\% of the full mosaic.  The exposure times per pointing are 7043 s, resulting in an average 5 $\sigma$ line sensitivity of $\sim$2$\times10^{-18}$ \fluxunit\ at a resolution of R$\sim$1600. The grism data of GOODS-N used here were obtained in February 2023.

The NIRCam/grism observations are reduced using the publicly available \texttt{grizli} code\footnote{\url{https://github.com/gbrammer/grizli}} (see also Brammer et al, in prep). Specifically, we start from the rate files that are obtained from the MAST archive. These are then aligned to a Gaia-matched reference frame. The direct images of a given visit are used to align the associated grism exposures. Before combination of the long-wavelength data, a custom bad-pixel map is applied, which masks residual bad pixels. 

Following \citet{Kashino22}, we use a median filtering technique to remove the continuum spectra for each row of the individual grism exposures. The filter uses a 12 pixel central gap, which avoids self-subtraction in case of strong emission lines. After the first pass filtering, pixels with significant line flux are identified and masked, before running the median filtering again.

For each source of interest from the photometric catalog (see next section), we derive an extraction kernel based on the image morphology in the F444W band and the segmentation map. This kernel is used to perform an optimal extraction of 1D spectra from the aligned and combined grism exposures. We use slightly modified sensitivity curves and spectral traces based on the v4 grism configuration files\footnote{\url{https://github.com/npirzkal/GRISMCONF}}.

In addition to the grism observations, FRESCO obtained imaging in the two short-wavelength filters F182M and F210M, as well as direct imaging in the long-wavelength filter F444W. The 5 $\sigma$ depths of these images are respectively 28.3, 28.1, and 28.2 mag, as measured in circular apertures of 0\farcs32 diameter.

We note that deeper NIRCam imaging data are available over $\sim$50\% of the FRESCO mosaics thanks to observations from the JADES team
\citep{Robertson2022ARAA}. However, these data are not yet publicly available for inclusion in this analysis.

\subsection{FRESCO Multi-Wavelength Catalogs\label{sec:catalog}}

In addition to the new JWST NIRCam data, we also make use of all ancillary HST imaging available in the GOODS-North field. Being a key extragalactic field for more than a decade, GOODS North has been targeted by all the main telescope facilities through a large number of programs.  A complete listing of HST programs can be found on the Hubble Legacy Field (HLF) release page\footnote{\url{https://archive.stsci.edu/prepds/hlf/}} (see also \citealt{Whitaker19} and \citealt{Illingworth16}). Most importantly, the field was covered with ACS imaging from GOODS \citep{Giavalisco2004}, with ACS and WFC3/IR imaging by the CANDELS survey \citep{Koekemoer11,Grogin11}, as well as WFC3/IR grism imaging by the AGHAST survey \citep{Weiner14}.

Here, we use a re-reduction of all available HST ACS and WFC3/IR data in the archive in filters that cover the FRESCO pointings, which we drizzled to a common pixel frame of 40 mas/pixel as the JWST NIRCam data. The 5 $\sigma$ depth (in the same 0\farcs32 diameter apertures) for the ancillary data are well-matched to the FRESCO NIRCam imaging: they range from $\sim$28.6 mag for the ACS data (F435W, F606W, F775W, F814W, and F850LP) to $\sim$28.2 mag for the WFC3/IR data (F105W, F125W, F160W), with the exception of F140W that reaches to $\sim27.4$ mag. 

In total, we derive photometry in 12 bands for the GOODS-North data set. The F444W images are used as a detection image, and we run SExtractor \citep{Bertin96} in dual image mode to measure matched-aperture photometry. The images are all PSF-matched to the F444W detection image, and fluxes are corrected to total using the default SExtractor AUTO parameters in addition to a small correction for remaining flux outside the AUTO aperture based on WebbPSF's F444W curve of growth \citep{Perrin14}.

\subsection{Spectroscopic Sample of H$\alpha$ Emitters\label{sec:zdist}}

We briefly summarize the construction of the H$\alpha$ catalog at $z\approx4.9-6.6$ over the GOODS-North FRESCO field and refer readers to Brammer et al. (2023, in prep) and Naidu et al. (2023, in prep) for more detailed description of the overall methods and catalog validation. Line extractions from the grism data are based on \texttt{EAZY} photometric redshifts ($z_{\rm{phot}}$).  For each object, we search for lines in a window around $z_{\rm{phot}}\pm0.05\times(1+z_{\rm{phot}})$. The photometric redshifts are derived from the PSF-matched photometry described above (\S\ref{sec:catalog}).  Additionally, we also use the default \texttt{grizli} photometric redshifts catalog that uses a stack of the FRESCO imaging (F182M+F210M+F444W) as a detection image, and makes two different choices in performing the photometry (e.g., with different source deblending parameters).  We use both sets of photometric redshifts to marginalize over such choices. In practice, the best-fit redshifts at $z_{\rm{phot}}>4.5$ have a small median difference of $|\Delta_{\rm{z}}|<0.1$ and in most cases result in the same line-candidates, but a fraction ($\approx25\%$) have $|\Delta_{\rm{z}}|>0.3$. 

Every line extracted with S/N$>5$ is visually inspected to verify that its morphology is consistent with the source and also to look for false positives (e.g., from broad PAH features in the foreground that slip through the median filtering). Physical properties (e.g., stellar masses) are derived by jointly fitting the photometry and line-fluxes with a non-parametric star-formation history and Chabrier IMF using the \texttt{Prospector} SED-fitting code \citep{Leja2017_Prospector}.  We refer readers to Naidu et al. (2023, in prep) for further details about the modeling choices.

\begin{figure}
    \includegraphics[width=\columnwidth]{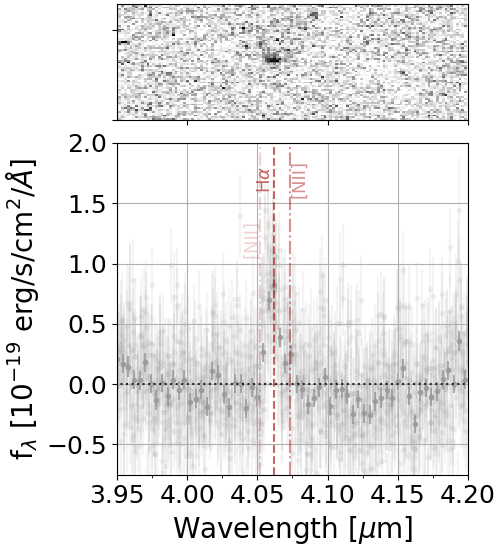}
    \caption{2D spectrum from the southern component to HDF850.1 (\textit{upper panel}) along with our 1D extraction from the direct image morphology (\textit{lower panel}).  The upper panel shows a zoomed-in 2D spectrum around the H$\alpha$ line after subtracting the continuum using a median-filtered technique following \citep[][: see \S\ref{sec:datadesc}]{Kashino22}.  The black dashed lines in the lower panel show the positions of the H$\alpha$ line (13$\sigma$ detection) and the [NII]$_{6585}$ line (4$\sigma$ detection) at the fitted redshift of $z=5.188$.  The expected wavelength of the [NII]$_{6549}$ line ($\approx$3$\times$ fainter than [NII]$_{6585}$: e.g. \citealt{Dojcinovic2023}) is also shown but does not show a significant detection in the FRESCO data. \label{fig:Spectrum}}
\end{figure}


\begin{figure}
\includegraphics[width=\columnwidth]{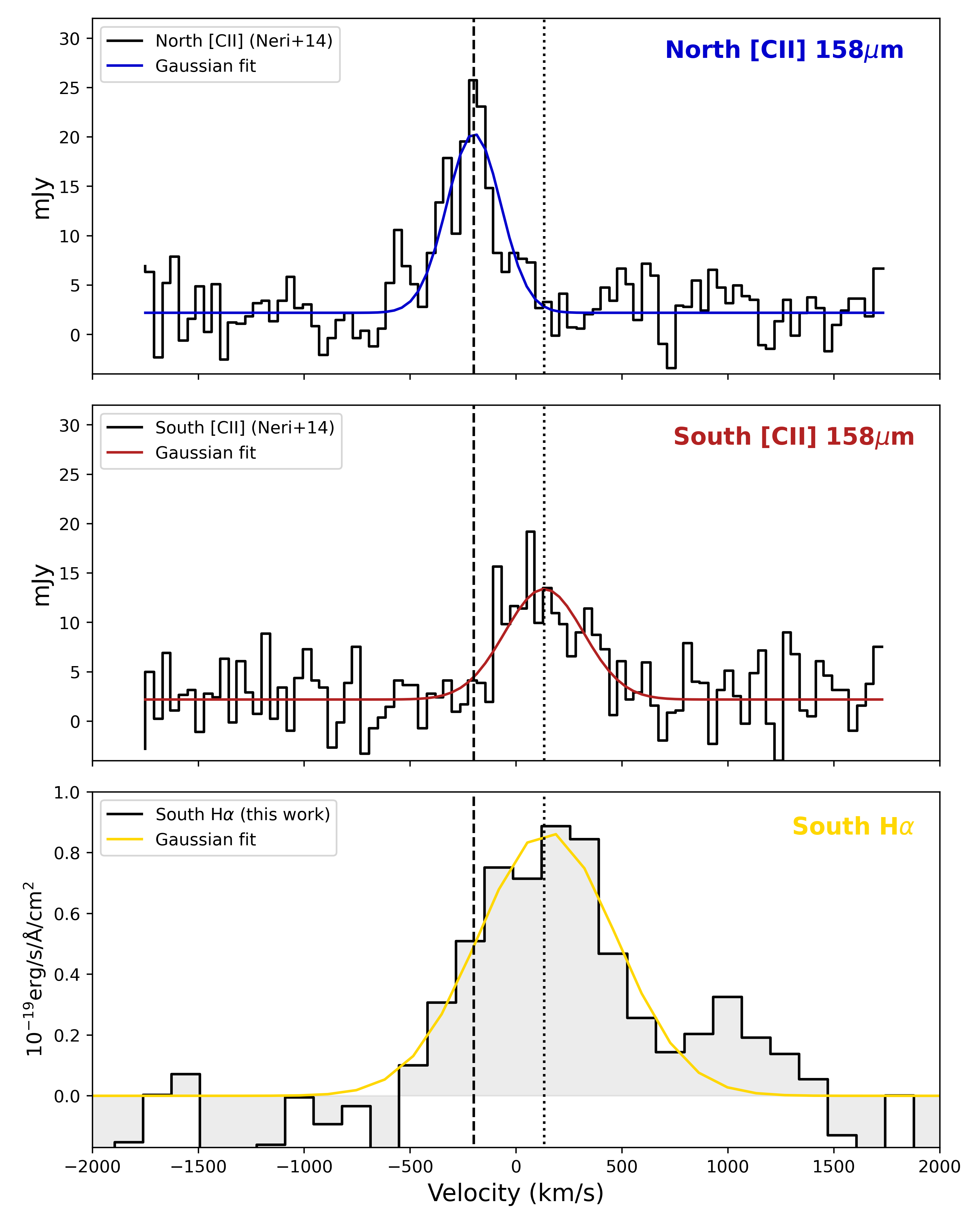}
\vspace{-0.3cm}
    \caption{Comparison between the [CII] detection from \citet{Neri2014} and our H$\alpha$ detection for both components. The upper two rows show the [CII] P$_{3/2} \rightarrow$P$_{1/2}$ line above the dust continuum from both component of HDF850.1 (\textit{black lines}) as well as the Gaussian fit to the lines (\textit{blue and red lines}) derived by \citet{Neri2014}.  The lowest row shows our detected H$\alpha$ line (\textit{black}) along with a Gaussian fit for the southern component (\textit{yellow}). Velocities for the [CII]$_{158\mu m}$ and H$\alpha$ lines are shown relative to the [CII] redshift determined by \citet{Neri2014} of z = 5.1853 (307.267 GHz). The dashed and dotted vertical lines are shown at the centroids of the Gaussian fits to  [CII] for the northern and southern components of HDF850.1, respectively,  to highlight the coinciding feature in the southern component. \label{fig:velcomp}}
\end{figure}

\section{Results}
\label{sec:results}

\subsection{Identification of H$\alpha$ Emission from HDF850.1}

One of the first sources we examined after creating catalogs of line emitting sources over the GOODS-North FRESCO field was the well-known dusty star-forming galaxy HDF850.1 with a spectroscopic redshift $z=5.1853$ from the 157.74$\mu$m [CII] line \citep{Walter2012,Neri2014}.  As we discussed in \S\ref{sec:intro}, that source evaded a direct redshift determination for $>$10 years following its initial discovery \citep[e.g.][]{Wagg2007}, and a redshift only came in 2012 thanks to a dedicated spectral scan for [CII] and various CO lines with PdBI \citep{Walter2012}.

The {\it JWST} F182M, F210M, and F444W NIRCam imaging we have available for HDF850.1 from FRESCO is shown in the center and right panels of Figure~\ref{fig:stamps} together with the blueshifted and redshifted high spatial resolution [CII] contours from PdBI.  Also shown is the imaging data of HDF850.1 at 0.8$\mu$m from F814W, again demonstrating that HDF850.1 radiates essentially no flux at $<$1$\mu$m.  From this imaging data, it is clear that while both components are detected in the NIRCam F444W data, only the southern component shows a clear detection in the F182M and F210M imaging data.  

We made use of the two prominent components of HDF850.1 in the F444W imaging data to construct segmentation maps, shown in the central and right panels of Figure~\ref{fig:stamps} as red and blue dashed lines.  We then used \texttt{grizli} to extract spectra of each component.  In Figure~\ref{fig:Spectrum}, we show both a two-dimensional and one-dimensional spectral extraction for the southern component to HDF850.1.  In the lower panel to Figure~\ref{fig:Spectrum}, we present a collapsed one-dimensional spectrum which not only reveals a $13\sigma$ detection of the H$\alpha$ line, but shows the detection of [NII]$_{6583}$ at $4 \sigma$ from HDF850.1.  No significant detection of [NII]$_{6548}$ is apparent in the FRESCO grism spectra, but this is not surprising given the fact that [NII]$_{6548}$ is $\approx$3$\times$ fainter than [NII]$_{6583}$ \citep[e.g.][]{Dojcinovic2023}.  

For H$\alpha$, we measure a total flux of $(6.4\pm0.5) \times 10^{-18}$ erg/s/cm$^2$ for the southern component to HDF850.1 and derive a redshift of 5.188$\pm$0.001. The [NII]/H$\alpha$ ratio we measure for the Southern component to HDF850.1 is $0.4 \pm 0.1$.  Such a high [NII]/H$\alpha$ ratio is commonly found for galaxies at higher masses with solar metallicities  \citep[e.g.][]{Shapley2015}.  Shocks could also be a contributing factor to the high [NII]/H$\alpha$ ratio we find \citep[e.g.][]{Kewley2013,Freeman2019} -- which would not be especially surprising given the apparent merging activity in HDF850.1 based on its two component structure \citep{Neri2014}.

\begin{figure}
\includegraphics[width=\columnwidth]{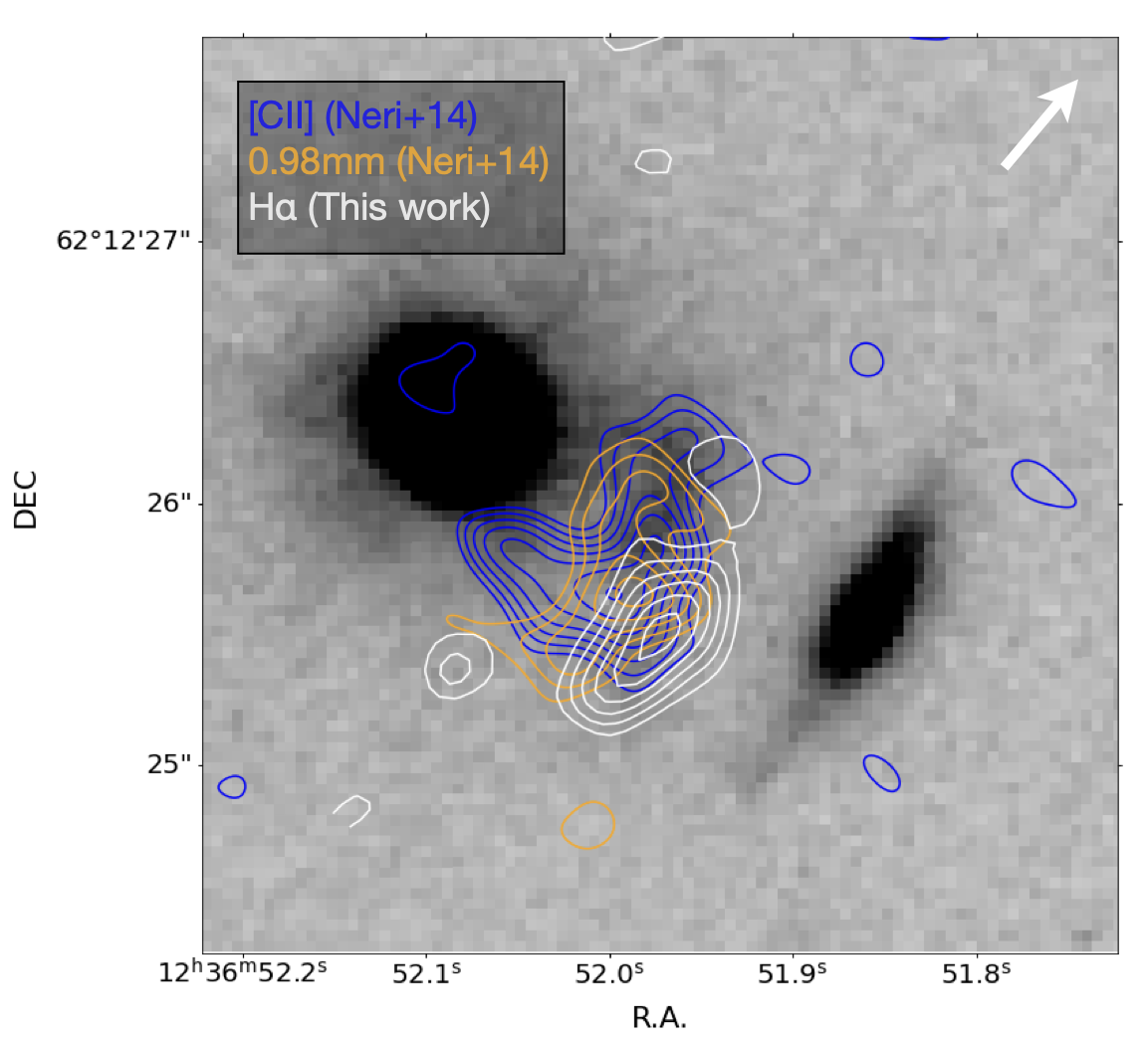}
\caption{Spatial distribution of line and dust emission relative to the {\it JWST} NIRCam F444W imaging observations of HDF850.1.  The extracted H$\alpha$ map is shown in white contours over the {\it JWST} NIRCam F444W imaging. The velocity averaged [CII]$_{158\mu{\rm m}}$ line detection and the 0.98mm dust continuum from \citet{Neri2014} are shown for comparison. All contours start at $3\sigma$ and increase in steps of $1\sigma$.\label{fig:Ha_map}}
\end{figure}

The redshift measurement we derived from the H$\alpha$ line implies a velocity offset of 133$\pm$34 km/s relative the systemic redshift measurement $z=5.1853$ derived by \citet{Neri2014}.  \citet{Neri2014} find a velocity offset of 130 km/s for the [CII] line from the southern component to HDF850.1, which is almost exactly the same velocity offset in [CII] for the southern component as we find here.  A detailed comparison of the line profiles and intensity of our new H$\alpha$ detection is shown in Figure~\ref{fig:velcomp} for the Southern component to HDF850.1, both in terms of the raw extraction (\textit{gray histogram}) and Gaussian fits to the lines (\textit{yellow lines}). Figure~\ref{fig:velcomp} also shows the line profiles for the northern and southern components to HDF850.1 as found by \citet{Neri2014}.  To further emphasize this similarity, a vertical dotted line is shown at the velocity offset for the southern component found by \citet{Neri2014}.  From Figure~\ref{fig:Ha_map}, it is furthermore clear that H$\alpha$ emission we detect (\textit{white contours}) appears to mostly originate from the southern component of HDF850.1, being offset from both the [CII] and continuum IR emission lying to the north (\textit{blue and orange contours, respectively}).  

Additionally, the H$\alpha$ line we extract from the southern component of HDF850.1 appears to be very broad overall.  Ignoring for the moment the impact spatial extension has on the width of the lines, the best-fit FWHM we find for the Southern component is of 7.6$\times$10$^2$ km/s.  To account for the impact that source morphology has in broadening the line, we make use of \texttt{grizli} to forward model the source based on its direct image morphology using the same dispersion direction as in the observations.  Based on this forward modeling, we find that the non-zero size of HDF850.1 contributes 6.2$\times$10$^2$ km/s to the measured FWHMs.  Subtracting this contribution in quadrature from the southern component, we derive a FWHM of (4.4 $\pm$ 0.9) $\times$ 10$^2$ km/s.  The width of this line is therefore consistent with what \citet{Neri2014} derive for [CII] from the southern component.  Given similar velocity offsets for both lines (130 km/s for [CII] vs. (1.3$\pm$0.3) $\times$ $10^2$ km/s for H$\alpha$), it seems clear that ISM material producing both lines show a very similar kinematic structure.

In contrast to the southern component, the northern component of HDF850.1 does not show a clear, localized detection of H$\alpha$ line emission.  As a result and due to the proximity of the two components and dispersion direction (shown in Figure \ref{fig:stamps}), line emission from the southern component partially extends into the same spatial region where H$\alpha$ line flux measurements need to be made for the northern component.  As a result of these challenges, we only report an upper limit to the H$\alpha$ flux for the northern component of $f_{\text{H}\alpha} < 2.1 \times 10^{-18}$ erg/s/cm$^2$ (3$\sigma$).  We derived this upper limit by comparing the observations with the spatial profile expected for H$\alpha$ emission from the northern component assuming a similar spatial distribution to the continuum light in the direct image.  By computing the 2D least squares residuals between the expected and observed line morphologies, we concluded that there is no significant H$\alpha$ line flux emanating from the northern component, and any line flux evident in the segmentation map for the northern component is consistent with contamination from the southern component. The flux measurements for the two components are provided in Table~\ref{tab:properties}.  

\subsection{Impact of Dust Obscuration on the H$\alpha$ Line Emission}

Thanks to our new measurements of the H$\alpha$ fluxes for both components of HDF850.1, we can compute an observed SFR for each component.  By comparing this SFR to the SFR implied by the respective [CII] luminosities, we can attempt to estimate the approximate dust obscuration of each component.  We use the conversion factor from \citet{Kennicutt2012}:\footnote{While we present SFRs and stellar masses using the \citet{Chabrier03} IMF and \citet{Kennicutt2012} presents their SFR relations using \citet{Kroupa2003} IMF, \citet{Kennicutt2012} emphasize that the relation for a \citet{Chabrier03} IMF is virtually identical.}
\begin{equation}
\label{eq:kennicutt}
{\rm SFR}_{{\rm H\alpha}} = L_{{\rm H\alpha}} (M_{\odot}/yr) / (10^{41.27}\,\textrm{erg/s}) 
\end{equation}
The SFR we estimate from the observed H$\alpha$ flux for the southern component is 6.0$\pm$0.5 (1.5/$\mu$) M$_{\odot}$/yr, while for the northern component the SFR we estimate is $<$1.8 (1.7/$\mu$) M$_{\odot}$/yr.  In specifying the SFR for each source, we have divided the result by the fiducial magnification factors for the two components derived in \citet{Neri2014} based on the isothermal model they constructed for a nearby $z=1.224$ elliptical galaxy, i.e., 1.7 for the northern component and 1.5 for the southern component.  Our quoted results in Table~\ref{tab:properties} include the factors (1.7/$\mu$) and (1.5/$\mu$) for clarity and to allow for a scaling of the results in case of updated lensing magnification factors.

We can estimate the approximate dust extinction for each component by relying on 
the measured [CII] luminosities for HDF850.1 from \citet{Neri2014} to estimate the total SFRs.  The ALPINE program \citep{LeFevre2020,Bethermin2020,Faisst2020} provided the following calibration of the $L_{[CII]}$-SFR relation \citep{Schaerer2020}:
\begin{equation}
{\rm SFR}_{{\rm [CII]}} = (L_{{\rm [CII]}}/10^{6.61} L_{\odot})^{0.855} (M_{\odot}/yr) 
\end{equation}
We infer [CII] SFRs of (1.6 $\pm$ 0.5)$\times$10$^2$ $(1.7/\mu)$ M$_{\odot}$/yr and (0.9 $\pm$ 0.3)$\times$10$^2$ $(1.5/\mu)$ M$_{\odot}$/yr for the northern and southern components to HDF850.1.  

Dividing the apparent H$\alpha$ SFRs by the [CII] SFRs, we infer SFR$_{\rm H\alpha}$/SFR$_{\rm [CII]}$ ratios of $<$0.01 and 0.07$\pm$0.03, respectively, demonstrating that even in the case of the clear detection of H$\alpha$ from the southern component of HDF850.1, dust attenuation would nevertheless appear to have a substantial impact on the observed line emission.  The SFR$_{\rm H\alpha}$/SFR$_{\rm [CII]}$'s we infer here are similar to what would be suggested by the $A_V$'s we derive for the two components in \S\ref{sec:sedmodel}, i.e., $\sim$0.02$_{-0.01}^{+0.04}$ and $\sim$0.12$_{-0.10}^{+0.25}$.

\begin{table}
\centering
\caption{Inferred characteristics of the two components of HDF850.1}
\vspace{-0.15cm}
\label{tab:properties}
\begin{tabular}{c|c} 
\hline \hline
\multicolumn{2}{c}{Northern Component}\\
\hline
Lensing Magnification $\mu$ & 1.7$^a$ \\
$L_{\rm [CII]}$ & 1.6$\times$10$^9$ $(1.7/\mu)$$^a$ $L_{\odot}$\\
SFR$_{\rm [CII]}$ & (1.6 $\pm$ 0.5)$\times$10$^2$ $(1.7/\mu)$$^a$ M$_{\odot}$/yr \\
$v_{\rm [CII]}$ & $-$200 km/s \\
FWHM$_{\rm [CII]}$ & 300 km/s \\
$f_{{\rm H\alpha}}$ & $< 2.1 \times 10^{-18}$ erg/s/cm$^2[3 \sigma]$ \\
SFR$_{\rm H\alpha}$ & $<$1.8 $(1.7/\mu)$ $M_{\odot}$/yr\\
SFR$_{\rm H\alpha}$ /
SFR$_{\rm [CII]}$ & $<$0.01 \\
$\tau_V$ & 3.9$_{-0.8}^{+0.6}$$^b$ \\
$\log_{10} M^* / M_{\odot}$ & 10.3$_{-0.3}^{+0.3}$ $+\log_{10}(1.7/\mu)$ $^b$ \\
$\log_{10} L_{IR}$ & 11.9$_{-0.2}^{+0.1}$ $+\log_{10}(1.7/\mu)$$^b$ \\
SFR$_{{\rm MAGPHYS}}$ & $64_{-17}^{+38}$ $(1.7/\mu)$$^b$ M$_{\odot}$/yr\\\hline

\multicolumn{2}{c}{Southern Component}\\
\hline 
Lensing Magnification $\mu$ & 1.5$^a$\\
$L_{\rm [CII]}$ & 0.8$\times$10$^9$ $(1.5/\mu)$$^a$ $L_{\odot}$\\
SFR$_{\rm [CII]}$ & (0.9 $\pm$ 0.3)$\times$10$^2$ $(1.5/\mu)$$^a$ M$_{\odot}$/yr \\
$v_{\rm [CII]}$ & 130$^a$ km/s \\
FWHM$_{\rm [CII]}$ & 410 km/s \\
$f_{{\rm H\alpha}}$ & $(6.5 \pm 0.4) \times 10^{-18}$ erg/s/cm$^2$ \\
$v_{\rm H\alpha}$ & (1.3 $\pm$ 0.3)$\times$10$^2$ km/s \\
FWHM$_{\rm H\alpha+[NII]}$ & (4.4 $\pm$ 0.9)$\times$10$^2$ km/s \\ 
SFR$_{\rm H\alpha}$ & $(6.5\pm 0.5)$ $(1.5/\mu)$ $M_{\odot}$/yr\\
SFR$_{\rm H\alpha}$ /
SFR$_{\rm [CII]}$ & 0.07$\pm$0.03 \\
$\tau_V$ & 2.1$_{-1.1}^{+1.6}$$^b$ \\
$\log_{10} M^* / M_{\odot}$ & 8.9$_{-0.1}^{+0.1}$ $+\log_{10}(1.5/\mu)$$^b$ \\
$\log_{10} L_{IR}$ & 10.3$_{-0.3}^{+0.4}$ $+\log_{10}(1.5/\mu)$$^b$ \\
SFR$_{{\rm MAGPHYS}}$ & $2.1_{-0.8}^{+3.3}$$^b$ $(1.5/\mu)$ M$_{\odot}$/yr\\
\hline
\end{tabular}
\begin{flushleft}
~\\\vspace{-0.4cm}
$^a$ From \citet{Neri2014}\\
$^b$ Derived using \textsc{MAGPHYS} (\S\ref{sec:sedmodel}: see Figure~\ref{fig:magphys})
\end{flushleft}
\end{table}

\begin{table}
\centering
\caption{Photometry Derived for HDF850.1}
\label{tab:photometry}
\begin{tabular}{c|c} 
\hline \hline
Band & Flux [AB mag]$^{a,b}$ \\
\hline
\multicolumn{2}{c}{Northern Component}\\
\hline
F435W & $>$29.0\\
F606W & $>$28.9\\
F775W & $>$29.2\\
F850LP & $>$28.5\\
F105W & $>$28.5\\
F125W & $>$28.3\\
F140W & $>$27.4\\
F160W & $>$28.2\\
F182M & $>$27.0\\
F210M & $>$27.0\\
F444W & 24.0$\pm$0.1\\\\
\multicolumn{2}{c}{Southern Component}\\
F435W & $>$29.0 \\
F606W & $>$28.9 \\
F775W & $>$29.2 \\
F850LP & $>$28.5 \\
F105W & $>$28.5\\
F125W & $>$28.3\\
F140W & $>$27.4\\
F160W & $>$28.2\\
F182M & 27.2$\pm$0.1\\
F210M & 27.2$\pm$0.1\\
F444W & 26.0$\pm$0.1\\
\hline 
\end{tabular}
\begin{flushleft}
$^a$ Derived using \textsc{galfit} \citep[][: \S\ref{sec:photometry}]{Peng2002} \\
$^b$ Upper limits are $2\sigma$ \\
\end{flushleft}
\end{table}

\begin{figure*}
\includegraphics[width=1.5\columnwidth]{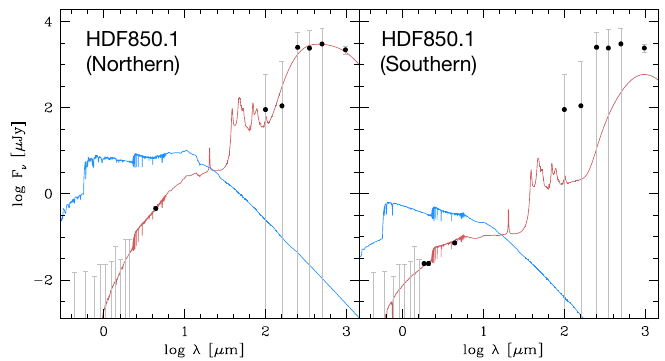}
\caption{Model spectral energy distribution derived by \textsc{MAGPHYS} to the northern and southern components of HDF850.1.  Blue and red lines show the emission from stars and from dust, respectively.  Optical and near-IR photometry shown with the solid circles (and 1$\sigma$ uncertainties) for both components are from Table~\ref{tab:photometry} and the far-IR constraints from Herschel as presented in \citet{Walter2012} and dust continuum at 0.9 mm from \citet{Neri2014}.  Since the spatial resolution of the Herschel observations ($>$15") is insufficient to provide independent measurements of the flux of each component of HDF850.1, we assume that the relative flux in each component is 50\%, similar to what is seen in the far-IR continuum from PdBI.  The procedure for performing the SED fits is detailed in \S\ref{sec:sedmodel}.\label{fig:magphys}}
\end{figure*}

\subsection{UV+Optical+far-IR SED Model of HDF850.1}

\subsubsection{Photometry on the Individual Components of HDF850.1\label{sec:photometry}}

Because of the very extended wings to the profile from a bright nearby elliptical galaxy, obtaining accurate flux measurements for HDF850.1 can be challenging to obtain using simple aperture photometry and therefore we elected to measure the flux for HUDF850.1 by modeling the light with analytic Sersic profiles using \textsc{galfit} \citep{Peng2002}.   This provides a very effective way of coping with contamination from the bright elliptical spilling over onto our aperture measurements.

We begin by using \texttt{galfit} to model the light in the NIRCam F444W band where both the northern and southern components of HDF850.1 are clearly detected, while also fitting to light in the nearby elliptical galaxy.  We model the light in the southern component of HDF850.1 with a single Sersic profile and light from the northern component as the sum of three different Sersic profiles.  After using these fits to measure the flux of both components to HDF850.1 in the F444W band, we fix the centers and shapes of the different contributors to both components and then refit the amplitudes in each passband.  

To account for the uncertainties associated with the extended wings of the nearby elliptical galaxy, we alternatively make use of two different Sersic parameters $n=2$ and $n=5$ in fitting for the contribution from that galaxy.   We take the flux uncertainty to be equal to the differences between the flux measurements in the two fits, and in cases where there is more than 0.4 mag differences between the flux measurements, we treat a component as undetected in a given passband.

We present the flux measurements we obtained for the two components of HDF850.1 in Table~\ref{tab:photometry}.   We measure a F444W-band magnitude of 24.0$\pm$0.1 mag for the northern component to HDF850.1 and 26.0$\pm$0.2 mag for the southern component.  We remark that the F444W flux we measure for the southern component is $\sim$4$\times$ what we would expect accounting for the line emission from H$\alpha$ alone, suggesting that stellar continuum from HDF850.1 contributes substantially to the F444W flux that we measure for the southern component.  For the northern component, the contribution from the stellar continuum seems to dominate the F444W flux given the absence of a clear line detection at its location. 

\subsubsection{SED Modeling of UV+Optical and far-IR Light from HDF850.1\label{sec:sedmodel}}

The JWST NIRcam observations presented here provide us with the first direct constraints on the stellar content of HDF850.1.  We complement the photometry listed in Table~\ref{tab:photometry} with the flux limits in the Herschel bands reported in \citet{Walter2012}, and the 1\,mm flux density continuum reported for the two components of the system in \citet{Neri2014}. We model the Spectral Energy Distribution of the northern and southern components of HDF\,850.1 using the high-redshift implementation of \textsf{MAGPHYS} \citep{daCunha2008,daCunha2015}. \textsf{MAGPHYS} assumes energy balance between energy absorbed by dust in the rest-frame optical range and re-emitted in the far-infrared. The stellar population is modeled based on \citet{Bruzual2003}'s spectral libraries, and assuming a delayed exponential function as star formation history. The best-fit model is shown in Fig.~\ref{fig:magphys}. We infer the best fit of the free parameters and their uncertainties from the distribution of the posterior probabilities, interpolated at the 16\%, 50\% and 84\% levels.  Our best-fit stellar masses, star formation rates, far-IR luminosities, and dust attenuation for the two components of HDF850.1 are provided in Table~\ref{tab:properties}.

The NIRCam observations sample the stellar continuum around the Balmer break, which lies at 2.1$\mu$m.  Our MAGPHYS fits suggest that both components of HDF850.1 are significantly dust reddened.  The northern component of HDF850.1 appears more massive and star forming, albeit with a similar specific SFR, compared to the southern component.  The precise flux measurements made possible thanks to our new NIRCam photometry pins down the stellar component of the fit; this however, combined with the energy balance assumption in the code, and the large uncertainties associated with the Herschel photometry of HDF850.1 as a whole, leads to the dust continuum emission in the southern component of HDF850.1 being underestimated.  A better sampling of the IR continuum emission separately for the two components of HDF850.1 is necessary to improve our characterization of the overall SED.

\begin{figure}
	\includegraphics[width=\columnwidth]{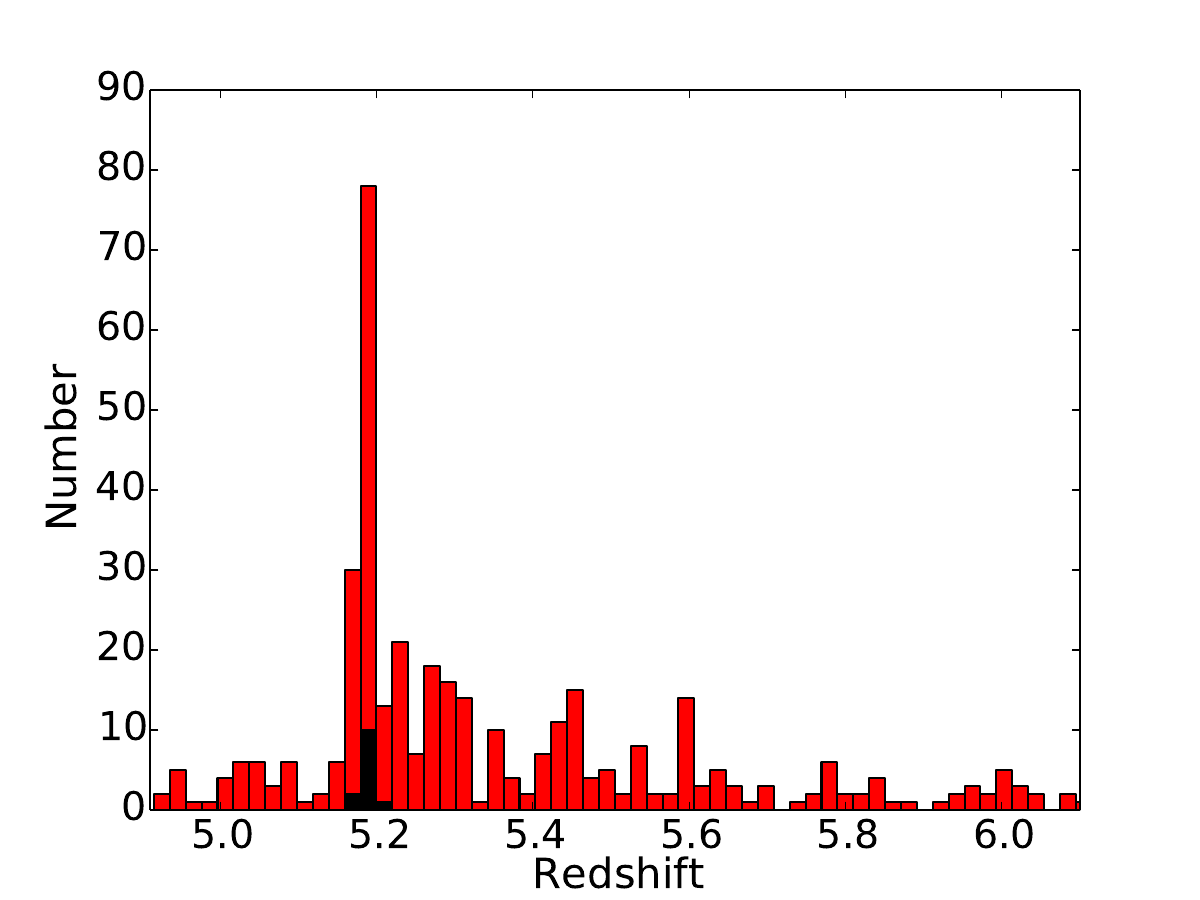}
 \vspace{-0.4cm}
    \caption{Redshift Distribution of Star-Forming Galaxies over the GOODS-North FRESCO field detected in H$\alpha$ at $>$5$\sigma$ (red histogram).  Sources shown in black in the histogram were identified earlier as part of the spectroscopic sample of \citet{Walter2012} and \citet{Calvi2021}.  Clearly, there is strong evidence for a substantial spike in the redshift distribution of galaxies at $z=5.17$-5.20, centered on the redshift of HUDF850.1.  Twenty four of the sources in this redshift spike were identified earlier as part of the \citet{AH2018} analysis leveraging the SHARDS data set.
    \label{fig:zdist}}
\end{figure}

\begin{figure*}
	\includegraphics[width=2\columnwidth]{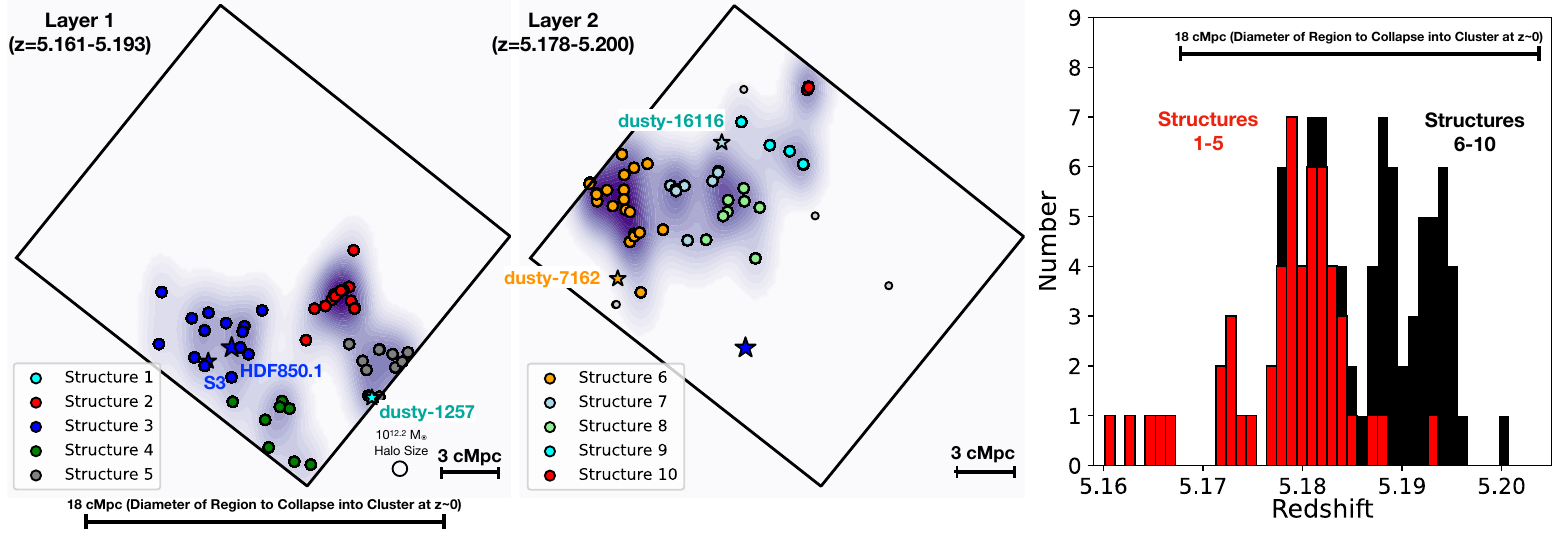}
    \caption{Spatial distribution of H$\alpha$ emitters (\textit{left and center panels}) in the prominent overdensity at $z=5.17$-5.20 in the GOODS-North FRESCO Field.  The black lines enclose the area covered by the FRESCO F444W grism observations.  Sources are shown in various color filled circles depending on the extended structure each was tentatively identified to belong based on the redshift and spatial position on the sky.  Extended structures 1, 2, 3, 4, and 5 are presented in the left panel, while structures 6, 7, 8, 9, and 10 are presented in the center panel.  The larger and smaller blue star show the position of HDF850.1 and another dusty star-forming galaxy S3 (see Figure~\ref{fig:2663}) which appears to be part of the same extended structure 3.  The orange, light blue, and cyan stars in the center panel show the positions of three dusty galaxies dusty-1257, dusty-7162, and dusty-16116 identified in extended structures 1, 6, and 7 by \citet{Xiao2023}.  The horizontal bars towards the lower region of the left and center panels  indicate the comoving radii (3 cMpc) of the most extended structures identified here and the comoving diameter (18 cMpc: \citealt{Chiang2017}) of the regions of the universe at $z\sim5.2$ that collapses into $>$10$^{14}$ M$_{\odot}$ galaxy clusters by $z\sim0$, respectively.  The small open circle indicates the expected size of 10$^{12.2}$ $M_{\odot}$ halos at $z\sim5.2$, The rightmost panel shows the number of sources as a function of redshift.  Sources in extended structures 1-5, shown as the red histogram, are distributed more towards the southern and western parts of the FRESCO GOODS-North field and mostly have redshifts between 5.165 and 5.185.  Meanwhile, sources in structures 6-10, shown as the black histogram, are distributed more towards the eastern and northern parts of the FRESCO field and mostly have redshifts between 5.185 and 5.196.  The horizontal bar shows the comoving length scale \citep{Chiang2017} of structures that collapse into $>$10$^{14}$ $M_{\odot}$ galaxy cluster by $z\sim0$.  The \citet{Chiang2017} suggest that all 10 extended structures identified here likely collapse into a single $>$10$^{14}$ $M_{\odot}$ galaxy cluster by $z\sim0$. \label{fig:lss}}
\end{figure*}

\begin{table*}
\centering
\caption{Extended structures Identified From $z=5.16$ to $z=5.31$ in the FRESCO GOODS North field$^a$}
\label{tab:structures}
\begin{tabular}{c|c|c|c|c|c|c|c|c|c|c|c}
\hline \hline
   & \multicolumn{4}{c}{Structure Center} & z$_{\rm mean}$ & $\sigma_v$ & Radius & & log$_{10}$ $\Sigma_i$ & \# of & Noteworthy\\
ID & RA  & Dec & $\Delta \alpha$ [cMpc] & $\Delta \delta$ [cMpc] & & [km/s] & [cMpc]$^b$ & $\delta_{g}$$^c$ & $M_{{\rm halo},i}$/$M_{\odot}$$^d$ & Members & Members\\\hline   
1 & 12:36:25.01 & 62:11:18.4 & 8.1 & $-$6.8 & 5.164 & 121 & 0.21 & ---  & 11.6 & 3 & Dusty-1257 \\
2 & 12:36:31.46 & 62:13:35.3 & 6.3 & $-$1.6 & 5.178 & 164 & 0.79 & 267  & 11.8 & 14 \\
3 & 12:36:55.00 & 62:12:39.7 & $-$0.0 & $-$3.7 & 5.180 & 176 & 2.83 & 30  & 12.6 & 16 & HDF850.1, S3\\
4 & 12:36:42.98 & 62:10:37.4 & 3.2 & $-$8.4 & 5.180 & 112 & 2.39 & 30  & 11.8 & 8 \\
5 & 12:36:23.06 & 62:12:06.9 & 8.6 & $-$5.0 & 5.184 & 211 & 0.74 & 33  & 11.5 & 9 & Dusty-7162\\
6 & 12:37:15.51 & 62:15:34.5 & $-$5.8 & 3.2 & 5.189 & 168 & 2.00 & 28  & 12.2 & 21 & Dusty-26715\\
7 & 12:37:01.42 & 62:16:20.0 & $-$1.8 & 4.7 & 5.194 & 54 & 0.95 & 80  & 11.2 & 8 \\
8 & 12:36:53.85 & 62:15:27.8 & 0.3 & 2.7 & 5.187 & 200 & 2.12 & 21  & 11.6 & 8 \\
9 & 12:36:44.98 & 62:16:58.4 & 2.7 & 6.2 & 5.190 & 163 & 0.43 & 127  & 11.4 & 5 \\
10 & 12:36:39.79 & 62:18:22.9 & 4.1 & 9.5 & 5.194 & 60 & 0.12 & ---  & 11.4 & 3 \\
11 & 12:37:03.90 & 62:12:26.5 & $-$2.4 & $-$4.2 & 5.220 & 182 & 2.23 & 21  & 12.1 & 9 & Dusty-4014\\
12 & 12:37:21.89 & 62:14:38.2 & $-$7.3 & 0.9 & 5.223 & 71 & 1.18 & 119  & 10.7 & 5 \\
13 & 12:36:34.76 & 62:16:56.2 & 5.5 & 6.3 & 5.268 & 275 & 1.49 & 12  & 11.1 & 5 \\
14 & 12:36:31.28 & 62:14:35.9 & 6.5 & 0.8 & 5.269 & 236 & 0.75 & 89  & 11.4 & 12 \\
15 & 12:36:26.99 & 62:14:24.3 & 7.7 & 0.4 & 5.301 & 165 & 0.66 & 54  & 12.5 & 8 & GN10\\
16 & 12:36:30.00 & 62:17:33.8 & 6.8 & 7.8 & 5.301 & 67 & 0.86 & ---  & 10.6 & 3 \\
17 & 12:37:19.53 & 62:15:38.7 & $-$6.8 & 3.3 & 5.302 & 211 & 2.09 & 8  & 11.6 & 4 \\
18 & 12:36:38.67 & 62:09:45.6 & 4.5 & $-$10.5 & 5.308 & 216 & 1.79 & ---  & 10.8 & 3 \\
\hline
\end{tabular}
\begin{flushleft}
  $^a$ These structures are presented in Figures~\ref{fig:lss} and \ref{fig:lss2}.  See \S\ref{sec:lss}\\
  $^b$ Estimated using Eq.~(\ref{eq:radius}).  Since the expected sizes of collapsed halos with masses of $10^{12.0}$ and $10^{12.2}$ $M_{\odot}$ are 0.3 cMpc and 0.4 cMpc, respectively, this suggests that the identified overdensities are composed of extended structure, rather than represent galaxies within a single halo.\\
  $^c$ This is the estimated amplitude of the overdensity of H$\alpha$ emitters in the cylindrical volume enclosed within the estimated radius of the structure and a $|\Delta z| < 2\sigma_v/c$ relative to H$\alpha$ emitters over the entire FRESCO GOODS-North volume.\\
$^d$ Four halos with mass $10^{12.0}$ $M_{\odot}$ and one halo mass with mass $10^{12.2}$ $M_{\odot}$ are expected within the FRESCO GOODS North $z=5.0$-6.0 volume.\\
\end{flushleft}
\end{table*}

\subsection{Extended Galaxy Structures Surrounding HDF850.1} 

\subsubsection{Redshift Overdensity at $z\sim5.2$ and Comparison with Earlier Studies}

Given the substantial clustering of other star-forming galaxies expected around massive galaxies like HDF850.1 and earlier results showing a significant overdensity of galaxies at $z\sim5.2$ \citep{Walter2012,AH2018,Calvi2021}, it is logical to make use of the substantial number of sources with spectroscopic redshifts in the GOODS North field from FRESCO to investigate this matter more extensively.

Using the techniques described in \S\ref{sec:zdist}, we constructed catalogs of H$\alpha$ emitters over GOODS North FRESCO field.  Figure~\ref{fig:zdist} shows  the number of sources as a function of redshift, and it is clear there is a huge overdensity of sources at $z=5.17$-5.20, with 100 sources found in that narrow redshift interval.  \citet{Walter2012} and \citet{Calvi2021} had both previously reported the same overdensity of galaxies at $z\sim5.2$, adding 13 and 6 spectroscopic members, respectively.  Analysis of the narrow-band SHARDS observations indicate 44 additional sources whose redshifts are consistent with lying in this overdensity \citep{AH2018}, but the redshift measurements from SHARDS are much less precise, i.e., $\Delta z\sim 0.07$, and so much more difficult to tie to distinct individual structures identified here.

Nineteen of the 100 H$\alpha$ emitters that we identified in the redshift range $z=5.16$ to $z=5.20$ were previously flagged as probable members of the $z\sim5.2$ overdensity by \citet{Walter2012}, \citet{AH2018}, and \citet{Calvi2021}.  These earlier studies appear to have been most efficient at identifying those member galaxies in the $z\sim5.2$ overdensity with the highest H$\alpha$ fluxes, likely as a result of these same sources showing more prominent Ly$\alpha$ emission lines for redshift determinations.   Three out of the seven brightest H$\alpha$ emitters (i.e., 43\%) were identified as part of earlier spectroscopic studies, as well as five out of the 14 brightest H$\alpha$ emitters (i.e., 36\%).   This compares with just 19 out of 100 sources (i.e. 19\%) that appeared in these earlier compilations.

Comparing our redshift measurements to those from \citet{Walter2012} and \citet{Calvi2021}, the redshift measurements we derive are $\Delta z = -0.009\pm0.011$ lower in the mean than those in the literature.  This is consistent with what we would expect for Ly$\alpha$ velocity offsets of 436$\pm$582 km/s, in the general range of what has been found in many studies at $z\sim3$-8 \citep[e.g.,][]{Erb2014,Tang2023}.  Somewhat unexpectedly, only six of the nineteen sources in the spectroscopic redshift catalogs of \citet{Walter2012} and \citet{Calvi2021} with coverage from FRESCO show up in our own catalogs of H$\alpha$ emitters.  Given the high completeness levels expected for our H$\alpha$-emitter searches, this suggests that previous spectroscopic samples were dominated by sources with high Ly$\alpha$ escape fractions.

There are three sources over the GOODS North field discussed by \citet{Walter2012} and \citet{Calvi2021} which have spectroscopic redshifts suggesting they are part of the prominent $z=5.17$--5.20 overdensity discussed here, but which lie outside the $\sim$62 arcmin$^2$ FRESCO mosaic.  These include a QSO at $z\sim5.18$ \citep{Barger2002} and two other galaxies  SHARDS20008777 and a source at 12:36:00.0, 62:12:26.1.  Nineteen of the plausible 50 sources with redshifts in the range $z=5.12$--5.27 from \citet{AH2018} lie outside the FRESCO mosaic.  Of the 31 sources from \citet{AH2018} which lie in the redshift range $z=5.12$--5.27 and lie within the FRESCO mosaic, 17 are members of our spectroscopic sample of H$\alpha$ emitters in the redshift range $z=5.16$ to $z=5.20$.

With 100 spectroscopic members of the overdensity at $z=5.17$--5.20, this overdensity is particularly extreme.  $\sim$28\% of the total number of $z=5.0$-6.0 H$\alpha$ emitters found over the HDF-North field lie in a $\Delta z\sim0.03$ interval.  This exceeds even the 45 galaxies present in $z\sim5.4$ overdensity identified over the GOODS-South field \citep{Helton2023} and also the 20 galaxy and 24 galaxy overdensities identified at $z\sim6.19$ and $z\sim6.33$, respectively, over the J0100+2802 field by \citet{Kashino22}.  

Given the much larger number of star-forming galaxies in the immediate vicinity of HDF850.1 than even present over the J0100+2802 field, this may suggest the halo masses associated with this overdensity may be even more extreme than the bright $z\sim6.33$ QSO J0100+2802 that the EIGER program targeted.  In particular, the $z=5.17$--5.20 overdensity extends over the entire GOODS-North FRESCO field, i.e., a 18 cMpc $\times$ 18 cMpc area and 15 cMpc along the line of sight, which translates to a 4.8$\times$10$^3$ cMpc$^3$ comoving volume.  Assuming a similar efficiency for line emission across the redshift interval $z=5.0$ to 6.0, this translates to an amplitude of $\delta_g + 1 = 8\pm1$ for the identified $z=5.17$--5.20 overdensity.  

Interestingly enough, the present overdensity is very similar in comoving size (17 cMpc $\times$ 20 cMpc $\times$ 26 cMpc and 17 cMpc $\times$ 20 cMpc $\times$ 36 cMpc), volume (6.7$\times$10$^3$ cMpc$^3$ and 1.2$\times$10$^4$ cMpc$^3$), and $\delta_g$ overdensity factors (4.8$\pm$1.8 and 9.5$\pm$2.0) to the overdensities at $z=3.065$ and $z=3.095$, respectively, identified over the SSA22 field by \citet{Steidel1998} and characterized in more detail by \citet{Topping2018}.

\subsubsection{Extended Galaxy Structures Around HDF850.1}
\label{sec:lss}

To gain insight into the extended structures that make up the overdensity at $z=5.17$--5.20, we constructed a list of candidate extended structures within the FRESCO GOODS-North volume.  For this, we counted the number of galaxies that lie closer than 1.2 cMpc to a given galaxy in the plane of the group and that lie within 10 cMpc along the line of sight, which corresponds to $\Delta z \sim 0.02$ (corresponding to a peculiar velocity of $\sim$1000 km/s).  We allowed for greater latitude in defining a galaxy structure along the line of sight, given the impact peculiar velocities have on the apparent position of a galaxy in three-dimension space.  This list was then ordered from galaxies with the largest number of neighbors in this volume to the smallest number of neighbors.   Sources identified as part of richer extended structures were then excluded from structures with a smaller number of members.  Finally, extended structures were then combined with adjacent extended structures if their mean centers in the plane of the sky were closer than 3.5 cMpc and their mean redshifts $\Delta z_{\rm mean}$ differed by 0.005.  

\begin{figure*}
	\includegraphics[width=2\columnwidth]{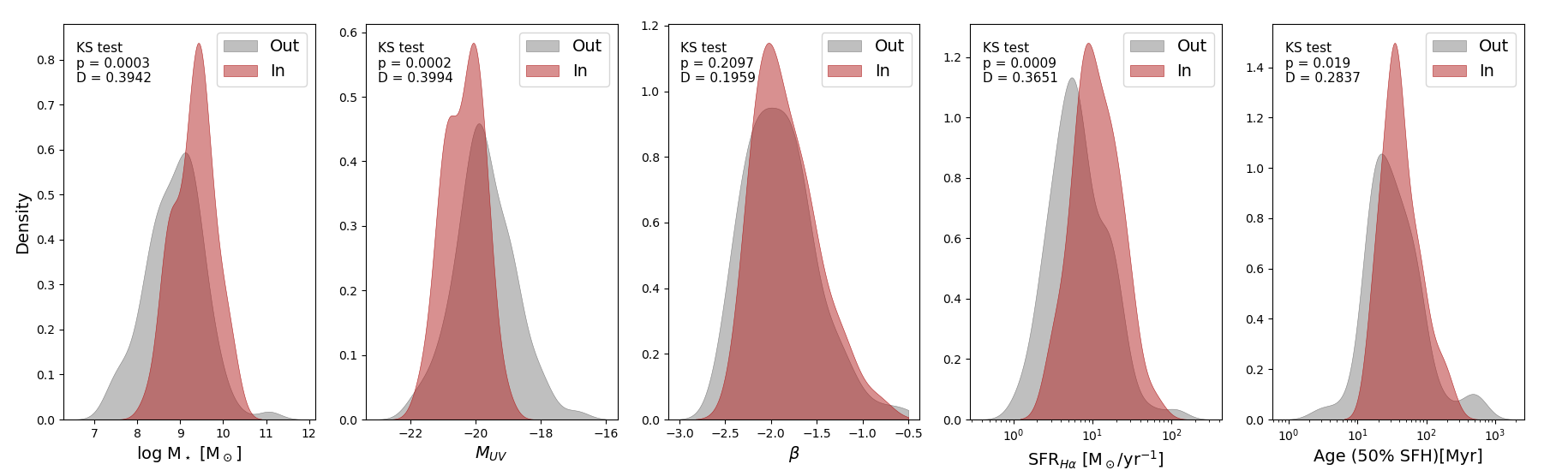}
    \caption{KDE Density plots showing the comparison of Prospector-inferred galaxy properties for sources inside and outside of the overdensity. The "in" sample (\textit{red}) contains 47 sources between redshifts $z=5.17$--5.20 in the overdensity and identified as part of extended structures, while the "out" sample (\textit{grey}) contains 86 sources at redshift $z=4.9$--5.5 excluding the overdensity redshift bin and the structures identified with the method detailed in section \ref{sec:lss}.  Shown in each panel are differences in the fractional cumulative distribution and associated probability that the two distributions are consistent using a Kolomogrov-Smirnov test.  The stellar masses, $UV$ luminosities, and star formation rates for galaxies inside the overdensities show a clear shift to higher values than galaxies outside the overdensities.
    \label{fig:fitshist}}
\end{figure*}

In determining whether to include sources in these extended structures, the $1\sigma$ dispersions in the redshifts for each structure are computed excluding 
sources that are $\Delta z\sim 0.01$ ($\approx$500 km/s) from the median redshift for a structure.  $\Delta z = 0.01$ is in excess of the redshift dispersion for all of the structures identified here and $\sim$2.9$\times$ the median 1$\sigma$ dispersion.  Any source that differs from the median redshift for a structure by 2.5$\sigma_{z}$ is excluded from a structure.

To help illustrate the 3-dimensional spatial distribution of galaxies and extended structures within this $z=5.17$--5.20 overdensity, we have included Figure~\ref{fig:lss}, which shows four different structures we have identified at $z=5.170$--5.185 (\textit{left panel}) and four different structures we have identified at $z=5.185$--5.200 (\textit{center panel}).  The rightmost panel shows the redshift distribution of sources in 
structures 1-4 (\textit{red histogram}) and structures 5-8 (\textit{black histogram}), pointing to the presence of a possible bi-modality 
in the $z=5.17$--5.20 overdensity.  Additionally, the structures located in the lower redshift mode lie predominantly to the south and western parts of the FRESCO field, while structures in the higher redshift mode, lie predominantly to the north and east.  

We also include several additional highly obscured galaxies identified by \citet{Xiao2023} within these overdensities with star-like symbols.  Interestingly enough, the same extended structure of galaxies that contains HDF850.1 also features another highly obscured galaxy S3.  Additionally, \citet{Xiao2023} report four other highly obscured galaxies in extended structures 1, 6, 7, 11, and 15.  The latter galaxy, previously named GN10, was identified as part of earlier work \citep{Daddi2009,Riechers2020}.

We have included a full list of the H$\alpha$ emitters  in the $z=5.17$--5.20 overdensity in Table~\ref{tab:sample} from Appendix \ref{sec:catappendix}.  Galaxies which lie in some nearby overdensities at $z\sim5.23$, $z\sim5.27$, and $z\sim5.3$ over the GOODS North FRESCO field are also presented in this table.  In Appendix~\ref{sec:structappendix}, we present the spatial distribution in the plane of the sky of these slightly higher redshift overdensities.  The spatial extent of the structures in these overdensities range from 1 cMpc to 4-5 cMpc in the plane of the sky, much less substantial than we see in the $z=5.17$-5.20 overdensity that contains HDF850.1, which is the focus of this manuscript.  

We include in Table~\ref{tab:structures} the central coordinates, mean redshift, nominal velocity dispersion, approximate radius, overdensity factor relative to the mean volume density of H$\alpha$ emitters in the GOODS North FRESCO field, and number of member galaxies of these extended structures.  The center of each structure is provided both in right ascension and declination and offset from the center of the FRESCO GOODS North field.  The median $1\sigma$ dispersion in redshift for the extended structures we identified is 0.0035, while the median velocity dispersion we find is 167$_{-3}^{+12}$ km/s.  This is only slightly ($\sim$18$\pm$8\%) lower than the velocity dispersions \citet{Kashino22} find, i.e., 212 km/s and 197 km/s, for the overdensities in the J0100+2802 field at $z\sim6.19$ and $z\sim6.33$, respectively.

Following \citet{Long2020} and \citet{Calvi2021}, we also use the stellar masses of the member galaxies to estimate the integrated halo masses of the extended structures that host the galaxies in the extended structures.  We do so by computing the total stellar mass we estimate from the member galaxies and then supposing an integrated star formation efficiency of 5\% based on detailed studies comparing the galaxy stellar mass functions to the halo mass functions \citep{Behroozi2018}, but we note that detailed studies \citep{Stefanon2021,McLeod2021} of the ratio of the stellar mass to halo mass density from galaxy stellar mass functions at $z\sim0$-10 integrated down to 10$^8$ $M_{\odot}$ give factors between 0.5\% to 2\%.  We make use of the masses from Naidu et al.\ (2023, in prep) SED fits with \texttt{Prospector} -- except for the sources in the optically-faint samples of \citet{Xiao2023}.  The derived integrated halo masses range from 10$^{10.6}$ $M_{\odot}$ and 10$^{12.6}$ $M_{\odot}$.  Not surprisingly, HDF850.1 and GN10 lie in the halos that belong to the structures with the highest integrated halo masses.  Assuming a survey area of $\sim$62 arcmin$^2$ and $\Delta z \sim 1$ volume ($\sim$2$\times$$10^{5}$ cMpc$^3$), we estimate that the GOODS North FRESCO volume should contain approximately one $10^{12.2}$ $M_{\odot}$ halo.  For this calculation, we made use of the public halo mass function calculator HMFcalc by \citet{Murray13} adopting a \citet{Planck18} cosmology.

The radius of each extended structure $R_{\rm structure}$ is computed using the standard formula from \citet{Heisler1985} used in computing dynamical masses of galaxy clusters based on a discrete number of galaxies with measured positions and line-of-sight velocities:
\begin{equation}
\label{eq:radius}
R_{\rm structure} = \frac{\pi N}{2\Sigma_{i<j} \frac{1}{R_{\perp ij}}}
\end{equation}
where $N$ is the number of galaxies in an extended structure and $R_{\perp ij}$ is the projected distance in the plane of the sky between galaxy $i$ and galaxy $j$. 

We also estimate overdensities for the extended structure we identified as part of Table~\ref{tab:structures} by comparing the volume density of H$\alpha$ emitters within the estimated radius and a $|\Delta z| < 2\sigma_z$ redshift width of an extended structure to that found between $z=5.0$ to 6.0 for the GOODS-North FRESCO field as a whole.  We compute overdensity factors between 8 and 267, far in excess of the linear regime.  Based on the volume density of H$\alpha$ emitters we find in the GOODS North FRESCO volume, i.e., 1.6$\times$10$^{-3}$ cMpc$^{-3}$, we can estimate a nominal bias for this population assuming abundance matching and find an average bias factor of $\sim$4.7 using the \citet{Trenti2008_cosmicvariance} "cosmic variance" calculator.  Adjusting for this bias factor, this is suggestive of overdensities in the matter distribution ranging from 2 to 57, which is significantly in excess of $\approx$1 for the linear regime but less than $200\times$ overdensities expected for completely collapsed structures.

It is interesting to ask whether we would expect protoclusters to be present over the FRESCO fields and indeed within the GOODS North FRESCO field.  Given that clusters generally have a mass of at least 10$^{14}$ $M_{\odot}$ \citep[e.g.][]{Kravtsov2012} at $z\sim0$, we can use abundance matching to estimate whether the FRESCO volume contains any such objects.  Using the halo mass functions from HMFcalc, we estimate the FRESCO GOODS-North volume to include 4 such clusters with mass 10$^{14.0}$ $M_{\odot}$ and one cluster reaching a mass of 10$^{14.4}$ $M_{\odot}$.

\subsubsection{Comparison of Structure Sizes to the Expected Sizes of Protoclusters}
\label{sec:sizes_protocluster}

We can compare the size of the extended structures we find to that expected from the detailed study of protocluster growth \citep{Chiang2017} making use of several semi-analytic galaxy formation models \citep{Guo2013,Henriques2015} applied to the Millennium simulations \citep{Springel2005}.  \citet{Chiang2017} present both the comoving sizes of collapsed protoclusters at various redshifts and the sizes of cosmic volumes that will collapse into $>$10$^{14}$ $M_{\odot}$ galaxy clusters by $z\sim0$.  Of relevance, \citet{Chiang2017} find protoclusters to extend to have a comoving radius of $\sim$9 cMpc at $z\sim5.2$, implying overdense structures to extend over a comoving distance of $\sim$18 cMpc, very similar to the spatial extent of the overdensities we have identified over the GOODS North FRESCO field (Figure~\ref{fig:lss}). \citet{Chiang2017} suggest that all 10 extended structures identified here likely collapse into a single $>$10$^{14}$ $M_{\odot}$ galaxy cluster by $z\sim0$.

We estimate that the halo masses of the most massive collapsed halo in those extended structures would be 10$^{12.0}$ $M_{\odot}$ and 10$^{12.2}$ $M_{\odot}$, respectively, at $z\sim5.2$.  Assuming that average overdensity factor in the collapsed halos is 200, the size of the halos would be 0.3 cMpc and 0.4 cMpc at $z\sim 5.2$, i.e., 8$"$ to 13$"$ in the plane of the sky.  In computing a radius of the identified structures, the above equation gives greater weight to galaxy pairs that have smaller separations.  The fact that many of these extended structures have computed sizes not especially larger than 0.4 cMpc suggests that at least a few of the member galaxies in the extended structures are part of the same halos.

\begin{figure*}
	\includegraphics[width=1.5\columnwidth]{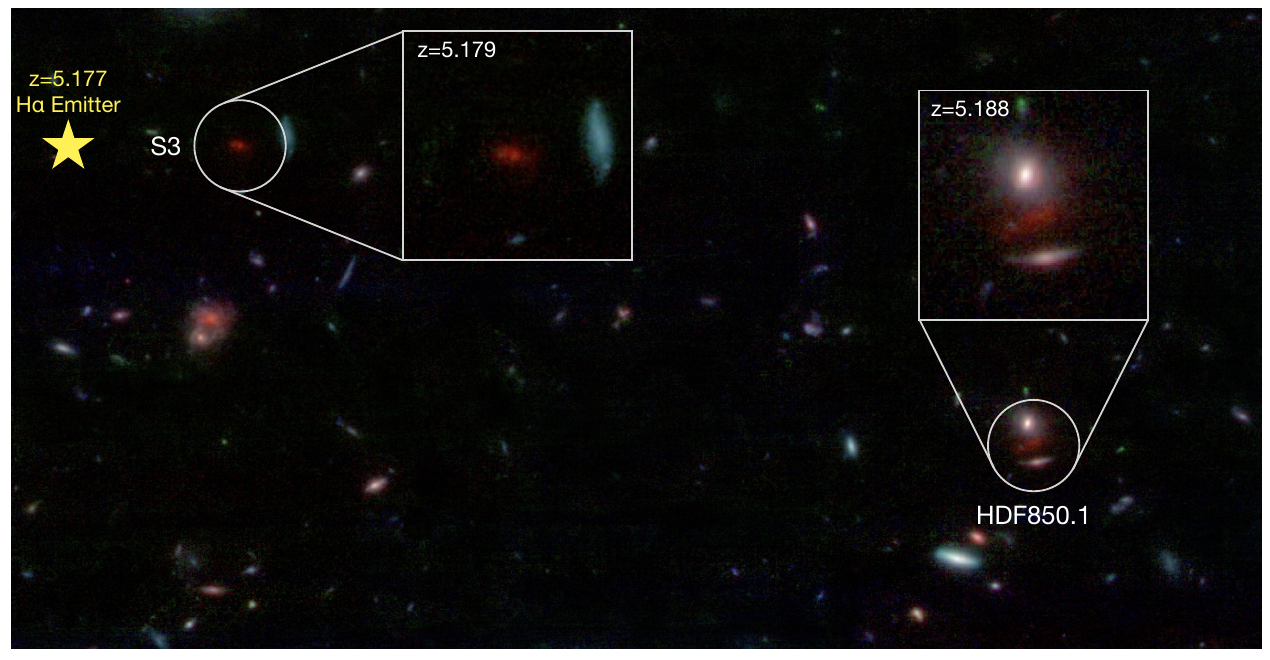}
    \caption{({\it upper panels}) Color composite image of another highly obscured star-forming source S3 at $z=5.179$ in the same extended structure \#3 as HDF850.1, as seen in a false color image (60$"$$\times$30$"$) made with the {\it JWST} NIRCam F182M, F210M, F444W data.  The redshift $z=5.179$ of this source is based on the detection of both H$\alpha$ ($>$20$\sigma$) and the [SII]$_{6719,6730}$ doublet ($\sim$9$\sigma$) in the FRESCO grism data.  Another H$\alpha$ emitter (ID 3946) at $z=5.177$ from the same extended structure (indicated with the yellow star) also lies within this same footprint and could reside in the same dark matter halo as S3 (being separated by 7$"$ [or 0.26 cMpc] in the plane of the sky).  Redshift measurements have been obtained for a large number of other highly obscured galaxies over the $\sim$124 arcmin$^2$ FRESCO mosaic using detected H$\alpha$ lines \citep{Xiao2023}, with typical S/N's of $>$10 for the sources with the highest inferred SFRs.  The detection of H$\alpha$ line emission from sources like HDF850.1 and S3 -- and many similarly dust-obscured and high-SFR sources -- shows the huge potential wide-area NIRCam grism observations have for mapping the build-up of massive, dust-enshrouded star forming galaxies in the early universe.
\label{fig:2663}}
\end{figure*}

\subsubsection{Characteristics of Galaxies in the $z=5.17$-5.20 Overdensity vs. Those Outside}
\label{sec:stellarpop_environment}

To put the extreme properties of HDF850.1 in context, we use the SED-fitting code Prospector \citep{Leja17,Johnson21} to derive galaxy properties from our extensive spectroscopically confirmed sample of galaxies in FRESCO GOODS-North. Our Prospector fits use FSPS with the MIST stellar models. We do not fit for redshift and set it to the derived $z_{\text{grism}}$ value infered by our pipeline using \texttt{grizli}. Along with the photometry, fluxes from emission lines identified in the Grism 1D spectra are also used as an input. We make use of \texttt{Prospector}'s more flexible non-parametric star-formation history (SFH) with 8 bins evenly separated in lookback time with a "continuity" prior for smoother SFH. The output catalog contains all sources detected in the detection image that were visiually inspected by at least one member of the FRESCO team along with the inferred properties from SED-fitting such as the retained stellar mass $M_\star$, UV slope $\beta$, star formation rates SFR or age at 50 \% of the total stellar mass. 

To highlight the distinct properties  galaxies may have within the overdensity or its structures and outside of it, we compare two samples. One contains all the galaxies in the $z = 5.17$--5.20 overdensity that have been identified as being part of a structure (\emph{"in"} with 40 sources) while the second sample contains all the remaining galaxies from $z = 4.9$ to $z = 5.5$ not identified in such structures (\emph{"out"} with 86) and also excluding the redshift bin $z = 5.17$--5.20. We also apply a magnitude cut of 27 AB mag in F182M. While this will limit our inferences to galaxies intrinsically brighter than $-20$ mag, this choice was made to alleviate the uncertainties on the inferred parameters from faint sources -- particularly for stellar masses -- and avoid a bias towards strong H$\alpha$ emitters only. We report the output of the \texttt{Prospector} fits for these two samples with kernel density estimate (KDE) plots as shown in Figure \ref{fig:fitshist}. We also compare SFRs inferred from H$\alpha$ using Eq. \ref{eq:kennicutt}. We use the H$\alpha$ flux corrected for dust attenuation with the empirical relation linking it to $\beta$ slopes from \citet{Shivaei2020}. Finally, we use a two-sample Kolmogorov–Smirnov (KS) test to comment on the dissimilarities in our selection of sources.\\

Based on the sizes of our two samples, we require a critical D-value of 0.3311 to demonstrate at 99.5\% confidence ($\alpha = 0.005$) that the distributions of the properties for sources inside and outside overdensities are different.   While we do not find a clear differences (at (99.5\% confidence) in the distributions for $\beta$ slopes and ages at 50\% of the SFH, the D-values and p-values for the masses and ages show clear differences between the two populations of galaxies at this redshift.  Our results suggest that the galaxies in the overdensity are more evolved, more massive and with higher SFRs than the galaxies outside of these structures.  The median stellar mass $M_\star$ in the overdensity is $\sim 2.5 \times 10^{9} M_{\odot}$ vs.   $\sim 8 \times 10^{8} M_{\odot}$ for the galaxies outside, a factor of $\sim$3 difference.  Earlier work by \citet{Steidel2005_protocluster} had shown that the masses and ages of galaxies in a similarly prominent (i.e, $\delta_g \sim 7$) overdensity at $z=2.3$ were twice as high as those outside the overdensity in line with our results.

More evidence from FRESCO for the most extreme sources being found in overdense environments can be found in the \citet{Xiao2023} H$\alpha$ selection of highly obscured sources over the GOODS North field.  In particular, the five $z=4.9$--6.6 sources with the highest estimated stellar masses and SFRs all lie within the overdensities at $z\sim5.16$--5.20 and $z\sim5.30$.


\section{Utility of Wide-Area Grism surveys for Mapping the Build-up of Dust-Obscured Galaxies at $z>5$}
\label{sec:potential}

The identification of a 13$\sigma$ H$\alpha$ emission line in the well-known dust-obscured galaxy HDF850.1 clearly demonstrates the enormous utility of the NIRCam grism for deriving redshifts for dust-obscured galaxies at $z>4$ and mapping out the build-up of these galaxies from early times.

Another particularly exciting example of such a source as relates to HDF850.1 is shown in Figure~\ref{fig:2663} and been given a source ID S3.  That source has a redshift $z=5.179$ derived from the detection of both H$\alpha$ ($>$20$\sigma$) and the [SII]$_{6719,6730}$ doublet ($\sim$9$\sigma$: \citealt{Xiao2023}).  Given the similar redshift and spatial proximity, S3 appears to be part of the same extended structure as HDF850.1.  Remarkably, this source shows even more extreme stellar masses and SFRs than HDF850.1 \citep{Xiao2023}.

Earlier, \citet{Oesch2023_FRESCO}  presented a third example of such a highly obscured source with a detected H$\alpha$ line in their Figure 10.  In parallel with the current study, \citet{Xiao2023} present a much larger sample of obscured, H$\alpha$ emitters at $z>5$, leveraging the FRESCO NIRCam grism data to derive specroscopic redshifts. 

The existence of large numbers of obscured galaxies with detectable H$\alpha$ emission may arise because of prevalent merging activity amongst the brightest far-IR sources \citep[e.g.][]{Clements1996,Tacconi2008} and the possibility that the merging activity could create less obscured sight lines by which H$\alpha$ photons could escape from galaxies \citep[e.g.,][]{LeReste2023}.  One of the components in a merging system might also be subject to substantially less dust obscuration, which would also enhance the detectability of IR luminous sources.  Note that many ultra-luminous far-IR bright galaxies at $z\sim1$-3 even reveal the presence of escaping Lyman-$\alpha$ photons \citep[e.g.][]{Chapman2003,Chapman2005} through sensitive rest-$UV$ spectroscopy.  Given the utility of the Ly$\alpha$ line, the ubiquity of H$\alpha$ line emission from $z>5$ ULIRGs is thus less surprising.

Of course, the non-detection of H$\alpha$ in the rather extreme star-forming source SPT0311-58 at $z=6.900$ with MIRI and the present non-detection of the northern component to HDF850.1 suggests that not every highly obscured source will be well detected in H$\alpha$ \citep{Alvarez2023_SPT0311} in NIRCam grism observations, but the detection of H$\alpha$ at high significance ($>$10$\sigma$) from the southern component to HDF850.1, GN10, and a handful of other far-IR luminous galaxies with the FRESCO grism observations \citep{Xiao2023} suggests that the selection of dusty star-forming galaxies could be moderately complete in very wide-area grism observations.

Given the identification of 7 highly obscured galaxies populating the $z=5.15$--5.32 structures over the GOODS North FRESCO field \citep{Xiao2023}, the extension of such a survey over much wider areas, e.g., $>$600 arcmin$^2$, has the potential to identify $>$70 obscured far-IR bright star-forming galaxies at $z>5$ while simultaneously allowing for a characterization of the structures in which the massive galaxies are forming.

Such a survey would yield even more far-IR luminous galaxies at $z>5$ than even the $\approx$42 (707 sources $\times$ $\approx$6\% at $z>4$) far-IR bright galaxies estimated to be identified at $z>4$ \citep{Dudzeviciute2020} over the 1 deg$^2$ AS2UDS program \citep{Stach2019}.  Such a program would yield many IR-bright sources at $z>4$ than the $\approx$4 far-IR bright galaxies identified over 184 arcmin$^2$ from the 2mm MORA program \citep{Casey2021_MORA}.  Even scaling a MORA-like program to 600 arcmin$^2$ would only result in a yield of $\approx$13 $z>4$ galaxies.

\section{Summary}
\label{sec:summary}

In this paper, we present the detection of a 13$\sigma$ $H\alpha$ line for HDF850.1 in the NIRCam F444W grism observations over the GOODS North field from FRESCO \citep{Oesch2023_FRESCO}, recovering a redshift of 5.188$\pm$0.001. Detection of H$\alpha$ in HDF850.1 is particularly noteworthy, given how obscured star formation from the source is.  HDF850.1 was one of the first submm galaxies to be identified with SCUBA \citep{Hughes1998} and evaded efforts to pin down its redshift for $>$10 years until the Plateau de Bure Interferometer secured the detection of [CII] and various CO lines from a spectral scan \citep{Walter2012}.  

In addition, HDF850.1 is clearly detected in the F444W imaging observations available over the source with NIRCam, with the emission segregated into a distinct northern and southern component.  These two distinct components were also evident in the earlier observations of [CII] from HDF850.1 with PdBI \citep{Neri2014}.  Modeling the SEDs of the two components of HDF850.1 based on the available HST + NIRCam F182M, F210M, and F444W imaging observations, we find a much higher SFR, stellar mass, and dust obscuration for the northern component than the southern component.  

The majority of the H$\alpha$ emission for HDF850.1 appears consistent with originating from the southern component, not only due to the spatial localization of the emission but also due to its showing a very similar $\Delta v$$\sim$130 km/s velocity offset to that seen in the southern component from [CII] emission \citep{Neri2014}.  Comparison of the SFR inferred from the observed H$\alpha$ emission with that seen from [CII] is suggestive of the H$\alpha$ emission from the southern component being 93$\pm$3\% attenuated and the northern component $>$98\% attenuated.

Leveraging redshift determinations possible from the FRESCO NIRCam grism observations in F444W over the GOODS North field, we note the existence of 100 galaxies in total in the $z=5.17-5.20$ interval, indicative of a huge 8$\times$ overdensity of galaxies in that redshift interval relative to the $z=5.0-6.0$ population of H$\alpha$ emitters we see over GOODS North.  Earlier work by \citet{Walter2012}, \citet{AH2018}, and \citet{Calvi2021} had previously demonstrated the existence of a substantial overdensity around HDF850.1, but with a much smaller number of spectroscopically confirmed sources than we find as part of this study.

Taking advantage of both the spatial and redshift  information we have of galaxies in the $z=5.17$-5.20 and slightly higher redshift ($z=5.22$--5.31) overdensities, we have organized sources into 18 extended structures and computed radii, masses, velocity dispersions for the structures.  The median $1\sigma$ dispersion in the identified structures in redshift and $v_{los}$ is 0.0035 and 167 km/s, which is only slightly (18$\pm$8\%) lower than what \citet{Kashino22} find for the extended structures in the J0100+2802 field at $z\sim6.19$ and $z\sim6.33$.

Interestingly, the comoving physical size of the
extended structures we find around HDF850.1 extend over a 18 cMpc $\times$ 18 cMpc $\times$ 15 cMpc volume, very similar to the $\sim$18 cMpc physical size \citet{Chiang2017} expect for protoclusters based on an analysis of various semi-analytic galaxy formation results building on the Millennium simulation (\citealt{Springel2005}: \S\ref{sec:sizes_protocluster}).  The link to the $z=5.17$-5.20 overdensity being the progenitor to a $>$10$^{14}$ $M_{\odot}$ $z\sim0$ galaxy cluster is further strengthened by noting that within the FRESCO GOODS North volume we would expect four $>$10$^{14}$ $M_{\odot}$ clusters and one $>$10$^{14.4}$ $M_{\odot}$ cluster to form by $z\sim0$.

Additionally, we have made a systematic comparison of galaxy masses, SFRs, $UV$ luminosities, ages, and apparent dust extinctions inside the $z=5.17$-5.20 overdensity to that outside the $z=5.17$-5.20 overdensity and other overdensities in the field.  We find strong evidence ($>$3$\sigma$) that galaxies inside the overdensities have higher stellar masses, SFRs, and $UV$ luminosities than those outside these overdensities.  This conclusion is strengthened by the inclusion of optically faint, massive, high-SFR dust-obscured galaxies at $z=5.0$-6.0, almost all of which lie inside some overdensity within the GOODS North.

In the future, it should be possible to significantly expand the number of dusty star-forming galaxies identified in the $z>5$ universe thanks to on-going  NIRCam grism observations from EIGER \citep{Kashino22}, ASPIRE \citep{Wang2023_ASPIRE}, a new NIRCam grism program from MIRI GTO team over GOODS South (program ID 4549), as well as deeper NIRCam grism observations approved over a deep NIRCam parallel field (program ID 4540) and various HFF clusters (program ID 2883, 3516, 3538).  Of course, to obtain the strongest constraints, a substantially wider area NIRCam program would be ideal, especially over areas that already have sensitive observations of the far-IR continuum as has been obtained by the ASPIRE program over clusters \citep{Wang2023_ASPIRE}, the ALMA GOODS program obtained over the CANDELS Deep region of the GOODS South \citep{Franco2020_GOODSALMA}, and ex-MORA program over the COSMOS field \citep{Casey2021_MORA}.

\section*{Acknowledgements}

We are grateful to Roberto Neri and collaborators for providing us with spatially resolved information on both the dust-continuum and [CII] line emission from their high spatial resolution PdBI observations. RJB acknowledges support from NWO grants 600.065.140.11N211 (vrij
competitie) and TOP grant TOP1.16.057.  The Cosmic Dawn Center (DAWN) is funded by the Danish National Research Foundation under grant No.\ 140.  Cloud-based data processing and file storage for this work is provided by the AWS Cloud Credits for Research program.  Support for this work was provided by NASA through grant JWST-GO-01895 awarded by the Space Telescope Science Institute, which is operated by the Association of Universities for Research in Astronomy, Inc., under NASA contract NAS 5-26555. RPN acknowledges funding from JWST programs GO-1933 and GO-2279. Support for this work was provided by NASA through the NASA Hubble Fellowship grant HST-HF2-51515.001-A awarded by the Space Telescope Science Institute, which is operated by the Association of Universities for Research in Astronomy, Incorporated, under NASA contract NAS5-26555. MS acknowledges support from the CIDEGENT/2021/059 grant, from project PID2019-109592GB-I00/AEI/10.13039/501100011033 from the Spanish Ministerio de Ciencia e Innovaci\'on - Agencia Estatal de Investigaci\'on. This study forms part of the Astrophysics and High Energy Physics programme and was supported by MCIN with funding from European Union NextGenerationEU (PRTR-C17.I1) and by Generalitat Valenciana under the project n. ASFAE/2022/025. RAM acknowledges support from the ERC Advanced Grant 740246 (Cosmic\_Gas) and the Swiss National Science Foundation through project grant 200020\_207349.

This work is based on observations made with the NASA/ESA/CSA James Webb Space Telescope. The data were obtained from the Mikulski Archive for Space Telescopes at the Space Telescope Science Institute, which is operated by the Association of Universities for Research in Astronomy, Inc., under NASA contract NAS 5-03127 for JWST. These observations are associated with program \# 1895.

This paper made use of several publicly available software packages. We are indebted to the respective authors for their work: \texttt{IPython} \citep{ipython},
    \texttt{matplotlib} \citep{matplotlib},
    \texttt{numpy} \citep{numpy},
    \texttt{scipy} \citep{scipy},
    \texttt{jupyter} \citep{jupyter},
    \texttt{Astropy}
    \citep{astropy1, astropy2},
    \texttt{grizli}
    (v1.7.11; \citealt{grizli,grizli2}),
    \texttt{EAZY} \citep[][]{Brammer08},
    \texttt{SExtractor} \citep[][]{Bertin96},

\section*{Data Availability}

All data used here are available from the Barbara A. Mikulski Archive for Space Telescopes (MAST: \url{https://mast.stsci.edu}), both in the form of raw and high level science products.

\bibliographystyle{mnras}
\bibliography{main} 

\appendix

\section{Extended Galaxy Structures at $z\sim5.22$-5.31}
\label{sec:structappendix}

\begin{figure*}
	\includegraphics[width=2\columnwidth]{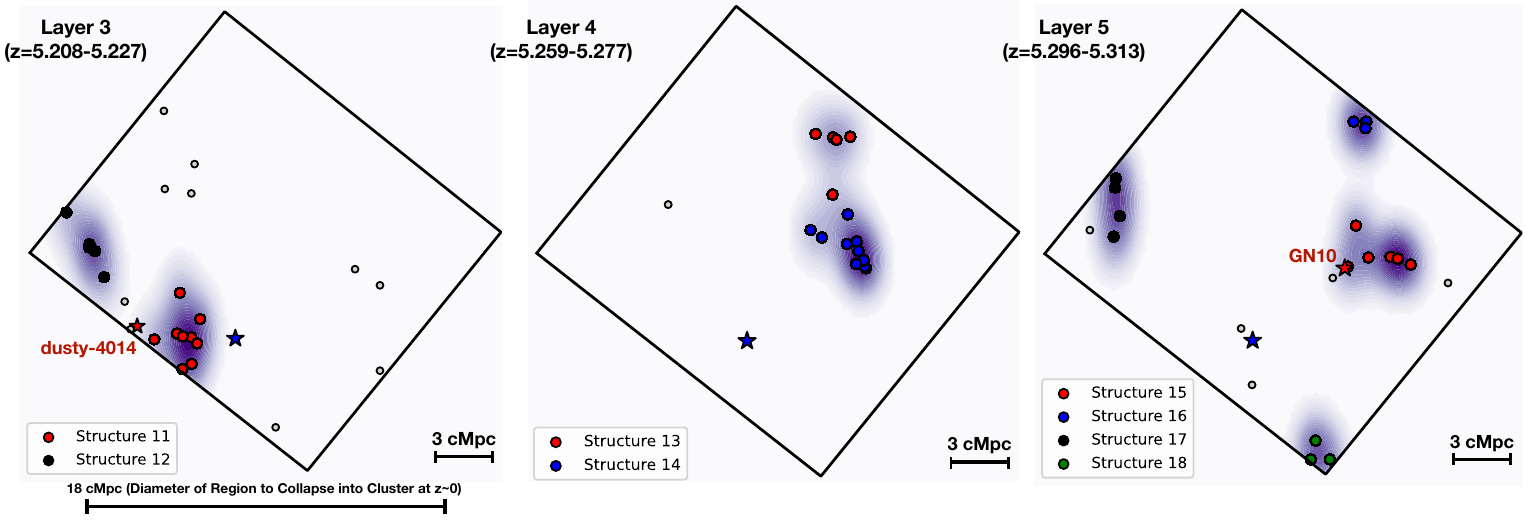}
    \caption{Similar to Figure~\ref{fig:lss} but for the identified structures in the overdensities of star-forming galaxies we have identified over the GOODS North FRESCO field at $z\sim 5.22$, $z\sim5.27$, and $z\sim5.30$.  The red star indicates the position of Dusty-4014 (dust-obscured galaxy identified by \citealt{Matthee2023_lrd} and \citealt{Xiao2023} showing H$\alpha$ in emission) and GN10 \citep{Daddi2009,Riechers2020}.  The spatial extent of the substructures range from 1 cMpc to 4 cMpc in size.\label{fig:lss2}}
\end{figure*}

For completeness and as an illustration of the extended galaxy structures that exist at redshift just above the prominent
$z=5.17$-5.20 overdensity.  Similar to Figure~\ref{fig:lss} from the main text, we show the spatial distribution of galaxies in 
the galaxy overdensities at the slightly higher redshifts of $z\sim5.23$, $z\sim5.27$, and $z\sim5.30$ for context.  As in the
main text, we have segregated sources in the overdensities into different structures using the same approach as described 
in \S\ref{sec:lss}.

While in some cases these overdensities consist of just a single group of $\sim$3-10 galaxies lying within 1 cMpc of each other (e.g., 
structures 9, 15), in other cases the overdensities extend over larger ($\sim$3-5 cMpc) scales.  However, even for the most extended structures in these overdensities, galaxies in these overdensities are much less spatially extended than in the prominent $z=5.17$-5.20 
overdensity that is the focus of this study.

\section{H$\alpha$ Emitters Identified in the $z\sim5.15$-5.31 Overdensity over the GOODS North FRESCO Field}
\label{sec:catappendix}

In this appendix, we provide a full catalog of the star-forming galaxies identified as part of the prominent $z=5.15$-5.31 overdensity over the GOODS North FRESCO field.  Coordinates for individual sources in the overdensity, together with their redshift determinations, flux measurements in the F210M band, H$\alpha$ fluxes, stellar masses, SFRs, $UV$-continuum slopes, and $V$-band dust attenuations are provided in Table~\ref{tab:sample}.  Also indicated are whether sources were previously reported as part of the \citet{Walter2012}, \citet{AH2018}, \citet{Riechers2020}, \citet{Calvi2021}, or \citet{Matthee2023_lrd} studies.  Given the $\Delta z \sim 0.07$ uncertainties in the photometric redshifts derived as part of \citet{AH2018}, sources in \citet{AH2018} are considered to be likely members of the overdensity at $z\sim5.2$ if their photometry redshifts are $z>5.13$ and $z<5.27$.

\begin{table*}
\centering
\caption{Catalogs of Star-Forming Galaxies at $z\sim5.15$-5.32 that are part of the overdensity in the FRESCO GOODS North Field.}
\label{tab:sample}
\begin{tabular}{c|c|c|c|c|c|c|c|c|c|c} \hline
     &    &     &            & $m_{210}$ & $f_{H\alpha}$\\
  ID & RA & DEC & $z_{{\rm spec}}$$^a$ & [mag] & [10$^{-18}$ ergs/s/cm$^2$] & log(M$_\star$) & $\beta$ & A$_V$ & SFR  &Ref$^b$ \\\hline\hline
\multicolumn{11}{c}{Extended Structure 1}\\
2152 & 12:36:25.39 & 62:11:19.8 & 5.161 & 25.6 & 15.0 $\pm$ 0.3 & 9.57$_{-0.49}^{+0.25}$ & $-1.27$$_{-0.11}^{+0.15}$ & $0.53$$_{-0.23}^{+0.14}$ & $31.5$$_{-15.4}^{+45.1}$ &  \\
2096 & 12:36:24.97 & 62:11:18.2 & 5.163 & 25.7 & 5.3 $\pm$ 0.4 & 9.79$_{-0.27}^{+0.18}$ & $-1.52$$_{-0.20}^{+0.23}$ & $0.50$$_{-0.17}^{+0.23}$ & $37.5$$_{-25.9}^{+72.0}$ &  \\
Dusty-1257$^b$ & 12:36:24.68 & 62:11:17.0 & 5.167 & 25.9 & 7.6 $\pm$ 0.3 & $>$9.7$^c$ & ---$^c$ & ---$^c$ & ---$^c$ & [6] \\
\\
  \multicolumn{11}{c}{Extended Structure 2}\\
8619 & 12:36:27.98 & 62:13:19.9 & 5.165 & 28.3 & 1.4 $\pm$ 0.2 & 8.40$_{-0.31}^{+0.38}$ & $-1.96$$_{-0.17}^{+0.27}$ & $0.81$$_{-0.26}^{+0.14}$ & $2.0$$_{-1.4}^{+2.9}$ &  \\
10338 & 12:36:30.69 & 62:13:44.0 & 5.172 & 28.4 & 1.2 $\pm$ 0.1 & 7.40$_{-0.13}^{+0.31}$ & $-2.38$$_{-0.11}^{+0.18}$ & $0.90$$_{-0.06}^{+0.04}$ & $0.5$$_{-0.1}^{+0.5}$ &  \\
10351 & 12:36:30.63 & 62:13:44.1 & 5.173 & 27.3 & 1.9 $\pm$ 0.2 & 8.65$_{-0.23}^{+0.26}$ & $-2.02$$_{-0.10}^{+0.10}$ & $0.69$$_{-0.22}^{+0.14}$ & $7.0$$_{-2.8}^{+5.2}$ &  \\
9967 & 12:36:31.37 & 62:13:39.0 & 5.173 & 27.8 & 2.4 $\pm$ 0.2 & 8.42$_{-0.68}^{+0.21}$ & $-2.15$$_{-0.14}^{+0.12}$ & $0.61$$_{-0.30}^{+0.24}$ & $3.0$$_{-2.1}^{+5.9}$ &  \\
9247 & 12:36:28.75 & 62:13:30.0 & 5.174 & 27.8 & 1.3 $\pm$ 0.2 & 8.36$_{-0.23}^{+0.61}$ & $-2.20$$_{-0.07}^{+0.08}$ & $0.84$$_{-0.48}^{+0.12}$ & $3.5$$_{-1.9}^{+6.1}$ &  \\
8799 & 12:36:33.73 & 62:13:22.9 & 5.175 & 27.2 & 1.7 $\pm$ 0.2 & 8.82$_{-0.30}^{+0.18}$ & $-2.02$$_{-0.09}^{+0.12}$ & $0.56$$_{-0.16}^{+0.26}$ & $8.2$$_{-4.4}^{+9.3}$ &  \\
9982 & 12:36:31.98 & 62:13:39.0 & 5.178 & 25.0 & 13.0 $\pm$ 0.4 & 10.07$_{-0.53}^{+0.17}$ & $-1.52$$_{-0.13}^{+0.12}$ & $0.57$$_{-0.15}^{+0.26}$ & $38.4$$_{-18.2}^{+31.6}$ &  \\
5848 & 12:36:37.52 & 62:12:36.0 & 5.179 & 26.1 & 5.5 $\pm$ 0.3 & 9.41$_{-0.15}^{+0.27}$ & $-1.69$$_{-0.09}^{+0.08}$ & $0.53$$_{-0.18}^{+0.26}$ & $27.3$$_{-23.6}^{+24.4}$ & [1,2] \\
8586 & 12:36:35.87 & 62:13:19.1 & 5.179 & 25.8 & 2.6 $\pm$ 0.4 & 9.44$_{-0.63}^{+0.25}$ & $-2.14$$_{-0.10}^{+0.10}$ & $0.70$$_{-0.25}^{+0.14}$ & $19.7$$_{-8.7}^{+34.3}$ &  \\
9754 & 12:36:31.83 & 62:13:36.2 & 5.179 & 26.5 & 4.7 $\pm$ 0.2 & 9.16$_{-0.32}^{+0.33}$ & $-2.02$$_{-0.08}^{+0.10}$ & $0.35$$_{-0.18}^{+0.28}$ & $27.9$$_{-14.5}^{+30.8}$ &  \\
10431 & 12:36:30.22 & 62:13:45.0 & 5.180 & 25.5 & 8.8 $\pm$ 0.3 & 9.65$_{-0.38}^{+0.18}$ & $-1.69$$_{-0.10}^{+0.09}$ & $0.56$$_{-0.14}^{+0.15}$ & $39.0$$_{-17.7}^{+37.9}$ &  \\
10723 & 12:36:29.11 & 62:13:49.0 & 5.181 & 27.6 & 1.5 $\pm$ 0.2 & 8.14$_{-0.21}^{+0.20}$ & $-2.46$$_{-0.08}^{+0.09}$ & $0.95$$_{-0.10}^{+0.03}$ & $2.5$$_{-0.9}^{+1.3}$ &  \\
15198 & 12:36:28.23 & 62:14:39.6 & 5.181 & 27.0 & 2.6 $\pm$ 0.2 & 8.96$_{-0.49}^{+0.17}$ & $-2.05$$_{-0.09}^{+0.08}$ & $0.64$$_{-0.22}^{+0.20}$ & $10.1$$_{-6.0}^{+14.4}$ &  \\
9362 & 12:36:32.49 & 62:13:31.3 & 5.181 & 25.6 & 8.9 $\pm$ 0.3 & 9.51$_{-0.26}^{+0.12}$ & $-2.15$$_{-0.06}^{+0.09}$ & $0.47$$_{-0.19}^{+0.19}$ & $60.6$$_{-25.6}^{+33.6}$ &  \\
8576 & 12:36:28.07 & 62:13:19.5 & 5.182 & 26.7 & 5.1 $\pm$ 0.2 & 9.11$_{-0.13}^{+0.16}$ & $-1.89$$_{-0.12}^{+0.10}$ & $0.51$$_{-0.11}^{+0.16}$ & $21.3$$_{-10.2}^{+11.2}$ &  \\
\\
\multicolumn{11}{c}{Extended Structure 3}\\
5578 & 12:37:06.17 & 62:12:30.7 & 5.172 & 27.8 & 1.3 $\pm$ 0.2 & 8.43$_{-0.24}^{+0.19}$ & $-2.06$$_{-0.12}^{+0.14}$ & $0.74$$_{-0.28}^{+0.16}$ & $4.2$$_{-1.7}^{+3.1}$ &  \\
7709 & 12:36:59.75 & 62:13:06.1 & 5.173 & 27.0 & 2.9 $\pm$ 0.3 & 8.87$_{-0.30}^{+0.20}$ & $-1.88$$_{-0.12}^{+0.12}$ & $0.58$$_{-0.11}^{+0.17}$ & $9.4$$_{-4.4}^{+7.2}$ &  \\
3946 & 12:36:57.24 & 62:12:00.8 & 5.177 & 26.7 & 2.0 $\pm$ 0.3 & 8.93$_{-0.43}^{+0.55}$ & $-1.55$$_{-0.12}^{+0.15}$ & $0.61$$_{-0.09}^{+0.10}$ & $7.0$$_{-3.6}^{+10.0}$ &  \\
6654 & 12:36:57.28 & 62:12:49.3 & 5.178 & 26.5 & 2.0 $\pm$ 0.3 & 8.87$_{-0.20}^{+0.31}$ & $-2.19$$_{-0.07}^{+0.08}$ & $0.89$$_{-0.14}^{+0.07}$ & $8.7$$_{-3.3}^{+6.8}$ & [2] \\
S3 & 12:36:56.56 & 62:12:07.4 & 5.179 & 28.2 & 5.3 $\pm$ 0.3 & $>$9.7$^{c}$ & ---$^{c}$ & ---$^{c}$ & ---$^{c}$ & [6]\\
5286 & 12:36:50.32 & 62:12:26.1 & 5.180 & 26.4 & 5.0 $\pm$ 0.3 & 8.78$_{-0.16}^{+0.20}$ & $-2.08$$_{-0.08}^{+0.12}$ & $0.70$$_{-0.13}^{+0.13}$ & $9.7$$_{-2.9}^{+5.0}$ &  \\
6557 & 12:36:49.86 & 62:12:47.7 & 5.180 & 25.8 & 7.9 $\pm$ 0.4 & 9.69$_{-0.61}^{+0.23}$ & $-1.48$$_{-0.11}^{+0.12}$ & $0.55$$_{-0.17}^{+0.25}$ & $27.4$$_{-15.4}^{+39.3}$ &  \\
8406 & 12:36:46.07 & 62:13:16.9 & 5.180 & 27.2 & 1.7 $\pm$ 0.2 & 8.37$_{-0.26}^{+0.19}$ & $-2.23$$_{-0.07}^{+0.08}$ & $0.94$$_{-0.09}^{+0.04}$ & $3.1$$_{-1.5}^{+1.5}$ & [2] \\
10218 & 12:37:05.64 & 62:13:42.1 & 5.181 & 26.3 & 3.4 $\pm$ 0.4 & 8.96$_{-0.26}^{+0.37}$ & $-2.21$$_{-0.06}^{+0.06}$ & $0.90$$_{-0.17}^{+0.07}$ & $8.5$$_{-2.2}^{+4.6}$ &  \\
3176 & 12:36:52.03 & 62:11:44.9 & 5.182 & 25.7 & 8.7 $\pm$ 0.5 & 9.20$_{-0.28}^{+0.41}$ & $-1.56$$_{-0.12}^{+0.14}$ & $0.65$$_{-0.13}^{+0.16}$ & $21.1$$_{-11.8}^{+14.6}$ &  \\
7003 & 12:36:49.51 & 62:12:55.2 & 5.182 & 26.1 & 3.9 $\pm$ 0.2 & 9.44$_{-0.50}^{+0.32}$ & $-1.88$$_{-0.07}^{+0.14}$ & $0.59$$_{-0.20}^{+0.15}$ & $17.9$$_{-9.1}^{+15.0}$ &  \\
7296 & 12:36:53.09 & 62:12:59.5 & 5.183 & 26.6 & 3.6 $\pm$ 0.4 & 8.65$_{-0.34}^{+0.54}$ & $-1.89$$_{-0.07}^{+0.08}$ & $0.79$$_{-0.12}^{+0.13}$ & $4.9$$_{-2.0}^{+5.2}$ & [2] \\
4528 & 12:36:59.25 & 62:12:12.3 & 5.184 & 26.7 & 3.7 $\pm$ 0.3 & 9.34$_{-0.52}^{+0.29}$ & $-1.16$$_{-0.18}^{+0.16}$ & $0.57$$_{-0.14}^{+0.14}$ & $11.4$$_{-10.1}^{+11.9}$ &  \\
8178 & 12:36:56.51 & 62:13:13.5 & 5.184 & 26.8 & 4.3 $\pm$ 0.3 & 8.75$_{-0.39}^{+0.20}$ & $-2.21$$_{-0.11}^{+0.10}$ & $0.80$$_{-0.32}^{+0.13}$ & $5.5$$_{-3.0}^{+4.6}$ & [2,4] \\
4783 & 12:36:48.71 & 62:12:16.8 & 5.185 & 24.1 & 38.0 $\pm$ 0.7 & 10.40$_{-0.25}^{+0.12}$ & $-0.71$$_{-0.09}^{+0.14}$ & $0.65$$_{-0.14}^{+0.10}$ & $107.0$$_{-43.4}^{+121.0}$ &  \\
HDF850.1 & 12:36:51.97 & 62:12:26.0 & 5.188 & 24.0 & 11.0 $\pm$ 0.7 & 10.4$_{-0.2}^{+0.3}$ & --- & --- & --- & [1,6]\\
\\
\multicolumn{11}{c}{Extended Structure 4}\\
1586 & 12:36:42.60 & 62:11:04.4 & 5.177 & 25.8 & 3.6 $\pm$ 0.4 & 9.47$_{-0.39}^{+0.25}$ & $-1.97$$_{-0.08}^{+0.10}$ & $0.79$$_{-0.33}^{+0.14}$ & $24.5$$_{-13.5}^{+13.9}$ &  \\
128 & 12:36:36.57 & 62:09:45.1 & 5.178 & 27.5 & 2.0 $\pm$ 0.2 & 8.00$_{-0.41}^{+0.46}$ & $-2.27$$_{-0.08}^{+0.10}$ & $0.93$$_{-0.06}^{+0.04}$ & $1.5$$_{-0.6}^{+1.7}$ &  \\
391 & 12:36:44.76 & 62:10:08.8 & 5.178 & --- & 2.2 $\pm$ 0.2 & 8.67$_{-0.35}^{+0.24}$ & $-1.72$$_{-0.18}^{+0.23}$ & $0.65$$_{-0.17}^{+0.14}$ & $4.9$$_{-2.8}^{+4.7}$ &  \\
169 & 12:36:39.79 & 62:09:49.1 & 5.179 & 25.2 & 20.0 $\pm$ 0.3 & 9.93$_{-0.61}^{+0.18}$ & $-1.60$$_{-0.10}^{+0.14}$ & $0.53$$_{-0.17}^{+0.18}$ & $36.2$$_{-17.3}^{+57.1}$ & [2] \\
999 & 12:36:45.47 & 62:10:46.5 & 5.181 & 27.9 & 4.8 $\pm$ 0.2 & 7.26$_{-0.12}^{+0.47}$ & $-2.22$$_{-0.10}^{+0.11}$ & $0.87$$_{-0.05}^{+0.05}$ & $0.4$$_{-0.1}^{+0.5}$ &  \\
1512 & 12:36:40.72 & 62:11:02.0 & 5.182 & 26.0 & 5.0 $\pm$ 0.4 & 8.86$_{-0.26}^{+0.41}$ & $-2.24$$_{-0.06}^{+0.06}$ & $0.95$$_{-0.05}^{+0.03}$ & $12.2$$_{-6.6}^{+6.1}$ &  \\
1820 & 12:36:51.79 & 62:11:11.3 & 5.182 & 26.0 & 12.0 $\pm$ 0.3 & 9.64$_{-0.28}^{+0.24}$ & $-1.56$$_{-0.10}^{+0.11}$ & $0.53$$_{-0.20}^{+0.13}$ & $19.6$$_{-10.1}^{+22.3}$ &  \\
1852 & 12:36:42.17 & 62:11:12.2 & 5.184 & 24.9 & 7.4 $\pm$ 0.6 & 10.19$_{-0.48}^{+0.24}$ & $-1.70$$_{-0.10}^{+0.10}$ & $0.56$$_{-0.34}^{+0.21}$ & $86.9$$_{-51.8}^{+307.9}$ &  \\
\hline\hline
\end{tabular}
\begin{flushleft}
$^a$ Typical $1\sigma$ spectroscopic redshift uncertainties are $\pm$0.001.\\
$^b$ 1 = \citet{Walter2012}, 2 = \citet{AH2018}, 3 = \citet{Riechers2020}, 4 = \citet{Calvi2021}, 5 = \citet{Matthee2023_lrd}, 6 = \citet{Xiao2023}\\
$^c$ Presented in \citet{Xiao2023}\\
\end{flushleft}
\end{table*}

\begin{table*}
\centering
\caption{Catalogs of Star-Forming Galaxies at $z\sim5.15$-5.32 that are part of the overdensity in the FRESCO GOODS North Field.}
\label{tab:sample2}
\begin{tabular}{c|c|c|c|c|c|c|c|c|c|c} \hline
     &    &     &            & $m_{210}$ & $f_{H\alpha}$\\
  ID & RA & DEC & $z_{{\rm spec}}$ & [mag] & [10$^{-18}$ ergs/s/cm$^2$] & log(M$_\star$) & $\beta$ & A$_V$ & SFR &Ref$^a$ \\\hline\hline
\multicolumn{11}{c}{Extended Structure 5}\\
4796 & 12:36:20.66 & 62:12:16.8 & 5.166 & 26.9 & 2.7 $\pm$ 0.2 & 8.85$_{-0.35}^{+0.18}$ & $-1.85$$_{-0.09}^{+0.19}$ & $0.71$$_{-0.25}^{+0.13}$ & $5.8$$_{-3.8}^{+6.2}$ &  \\
3678 & 12:36:25.72 & 62:11:55.1 & 5.179 & 27.6 & 1.8 $\pm$ 0.3 & 8.33$_{-0.32}^{+0.25}$ & $-2.12$$_{-0.13}^{+0.16}$ & $0.79$$_{-0.17}^{+0.11}$ & $2.4$$_{-1.3}^{+2.5}$ &  \\
5143 & 12:36:23.18 & 62:12:23.3 & 5.179 & 27.6 & 3.5 $\pm$ 0.2 & 8.15$_{-0.39}^{+0.33}$ & $-1.49$$_{-0.17}^{+0.23}$ & $0.57$$_{-0.10}^{+0.16}$ & $2.4$$_{-1.4}^{+2.4}$ &  \\
2087 & 12:36:23.22 & 62:11:18.3 & 5.181 & 26.1 & 6.1 $\pm$ 0.5 & 9.32$_{-0.61}^{+0.59}$ & $-1.76$$_{-0.14}^{+0.14}$ & $0.45$$_{-0.26}^{+0.33}$ & $18.3$$_{-10.4}^{+31.9}$ &  \\
5587 & 12:36:29.12 & 62:12:30.7 & 5.182 & 27.1 & 5.2 $\pm$ 0.3 & 8.62$_{-0.35}^{+0.26}$ & $-1.84$$_{-0.16}^{+0.21}$ & $0.59$$_{-0.21}^{+0.12}$ & $4.7$$_{-2.4}^{+5.9}$ &  \\
3888 & 12:36:20.13 & 62:11:58.6 & 5.183 & 25.2 & 24.0 $\pm$ 0.5 & 9.72$_{-0.39}^{+0.30}$ & $-1.40$$_{-0.09}^{+0.12}$ & $0.45$$_{-0.12}^{+0.19}$ & $48.8$$_{-20.3}^{+54.3}$ &  \\
4251 & 12:36:18.77 & 62:12:06.3 & 5.183 & 27.9 & 1.5 $\pm$ 0.2 & 8.45$_{-0.43}^{+0.20}$ & $-2.04$$_{-0.13}^{+0.16}$ & $0.68$$_{-0.13}^{+0.15}$ & $3.5$$_{-1.8}^{+3.8}$ &  \\
5146 & 12:36:23.28 & 62:12:23.0 & 5.187 & 26.3 & 6.7 $\pm$ 0.3 & 9.30$_{-0.31}^{+0.20}$ & $-1.40$$_{-0.19}^{+0.15}$ & $0.52$$_{-0.14}^{+0.17}$ & $18.2$$_{-10.1}^{+26.5}$ &  \\
4919 & 12:36:17.66 & 62:12:19.3 & 5.188 & 27.8 & 1.7 $\pm$ 0.2 & 8.46$_{-0.36}^{+0.26}$ & $-1.95$$_{-0.23}^{+0.16}$ & $0.83$$_{-0.13}^{+0.09}$ & $1.3$$_{-0.6}^{+1.2}$ &  \\
4306 & 12:36:26.48 & 62:12:07.4 & 5.193 & 26.1 & 5.1 $\pm$ 0.3 & 9.55$_{-0.25}^{+0.20}$ & $-1.67$$_{-0.10}^{+0.11}$ & $0.52$$_{-0.25}^{+0.16}$ & $29.9$$_{-17.8}^{+43.0}$ & [1,2] \\
\\
\multicolumn{11}{c}{Extended Structure 6}\\
17877 & 12:37:08.08 & 62:15:08.2 & 5.178 & 28.0 & 1.9 $\pm$ 0.2 & 8.56$_{-0.30}^{+0.48}$ & $-2.02$$_{-0.12}^{+0.12}$ & $0.69$$_{-0.41}^{+0.15}$ & $3.4$$_{-2.0}^{+8.5}$ &  \\
10187 & 12:37:12.39 & 62:13:42.1 & 5.181 & 27.3 & 1.5 $\pm$ 0.3 & 8.31$_{-0.64}^{+0.38}$ & $-2.13$$_{-0.13}^{+0.12}$ & $0.88$$_{-0.13}^{+0.07}$ & $1.6$$_{-0.7}^{+2.0}$ &  \\
23929 & 12:37:15.63 & 62:16:23.5 & 5.184 & 25.7 & 5.6 $\pm$ 0.3 & 9.59$_{-0.53}^{+0.24}$ & $-2.15$$_{-0.06}^{+0.08}$ & $0.65$$_{-0.35}^{+0.24}$ & $25.7$$_{-9.9}^{+85.6}$ & [1,2,4] \\
17512 & 12:37:12.62 & 62:15:04.1 & 5.185 & 25.4 & 4.4 $\pm$ 0.3 & 9.95$_{-0.15}^{+0.10}$ & $-1.60$$_{-0.09}^{+0.09}$ & $0.72$$_{-0.22}^{+0.14}$ & $17.9$$_{-13.6}^{+30.8}$ &  \\
20673 & 12:37:17.87 & 62:15:40.7 & 5.187 & 26.7 & 1.8 $\pm$ 0.2 & 9.28$_{-0.26}^{+0.22}$ & $-1.96$$_{-0.08}^{+0.08}$ & $0.50$$_{-0.27}^{+0.22}$ & $16.9$$_{-10.7}^{+34.9}$ &  \\
22471 & 12:37:15.70 & 62:16:03.0 & 5.187 & 25.1 & 8.1 $\pm$ 0.4 & 10.23$_{-0.58}^{+0.21}$ & $-0.84$$_{-0.19}^{+0.12}$ & $0.52$$_{-0.18}^{+0.14}$ & $35.9$$_{-11.9}^{+62.3}$ &  \\
19988 & 12:37:14.51 & 62:15:32.5 & 5.188 & 25.3 & 4.7 $\pm$ 0.3 & 9.37$_{-0.32}^{+0.33}$ & $-2.24$$_{-0.08}^{+0.08}$ & $0.94$$_{-0.07}^{+0.04}$ & $26.3$$_{-11.3}^{+17.5}$ & [2,4] \\
21948 & 12:37:21.03 & 62:15:56.6 & 5.188 & 28.3 & 1.0 $\pm$ 0.2 & 8.04$_{-0.26}^{+0.18}$ & $-2.15$$_{-0.10}^{+0.08}$ & $0.77$$_{-0.17}^{+0.08}$ & $2.1$$_{-0.8}^{+1.1}$ &  \\
23257 & 12:37:22.39 & 62:16:11.3 & 5.188 & 26.1 & 5.1 $\pm$ 0.3 & 9.25$_{-0.16}^{+0.18}$ & $-2.03$$_{-0.09}^{+0.09}$ & $0.59$$_{-0.20}^{+0.21}$ & $28.8$$_{-14.3}^{+16.0}$ &  \\
16991 & 12:37:13.69 & 62:14:59.1 & 5.189 & 26.6 & 4.5 $\pm$ 0.4 & 8.76$_{-0.36}^{+0.32}$ & $-1.89$$_{-0.10}^{+0.08}$ & $0.66$$_{-0.14}^{+0.14}$ & $8.0$$_{-3.5}^{+4.9}$ &  \\
Dusty-7162 & 12:37:16.90 & 62:14:00.9 & 5.189 & 25.1 & 30.0 $\pm$ 0.6 & $>$9.8$^c$ & ---$^c$ & ---$^c$ & ---$^c$ & [6] \\
17347 & 12:37:13.70 & 62:15:02.5 & 5.189 & 28.1 & 2.2 $\pm$ 0.3 & 8.26$_{-0.43}^{+0.38}$ & $-1.30$$_{-0.27}^{+0.28}$ & $0.56$$_{-0.15}^{+0.18}$ & $1.9$$_{-0.9}^{+2.3}$ &  \\
20278 & 12:37:15.33 & 62:15:35.5 & 5.189 & 25.9 & 6.8 $\pm$ 0.3 & 9.86$_{-0.17}^{+0.15}$ & $-1.29$$_{-0.13}^{+0.10}$ & $0.47$$_{-0.13}^{+0.12}$ & $23.0$$_{-12.5}^{+21.8}$ &  \\
21183 & 12:37:20.93 & 62:15:47.2 & 5.189 & 27.0 & 2.7 $\pm$ 0.3 & 8.37$_{-0.34}^{+0.50}$ & $-2.16$$_{-0.10}^{+0.12}$ & $0.84$$_{-0.10}^{+0.09}$ & $2.9$$_{-1.3}^{+7.0}$ &  \\
23256 & 12:37:22.28 & 62:16:13.2 & 5.191 & 25.2 & 5.6 $\pm$ 0.5 & 9.74$_{-0.45}^{+0.41}$ & $-1.67$$_{-0.12}^{+0.10}$ & $0.73$$_{-0.27}^{+0.14}$ & $54.7$$_{-24.1}^{+104.4}$ &  \\
16264 & 12:37:14.53 & 62:14:51.5 & 5.192 & 26.1 & 8.2 $\pm$ 0.4 & 9.59$_{-0.16}^{+0.16}$ & $-1.67$$_{-0.10}^{+0.08}$ & $0.50$$_{-0.12}^{+0.12}$ & $21.7$$_{-10.2}^{+20.5}$ &  \\
22058 & 12:37:20.97 & 62:15:58.0 & 5.192 & 27.1 & 4.1 $\pm$ 0.2 & 8.72$_{-0.31}^{+0.26}$ & $-2.04$$_{-0.09}^{+0.10}$ & $0.53$$_{-0.16}^{+0.22}$ & $8.7$$_{-4.5}^{+11.6}$ &  \\
22414 & 12:37:18.99 & 62:16:02.7 & 5.193 & 26.9 & 3.4 $\pm$ 0.3 & 8.65$_{-0.36}^{+0.38}$ & $-2.01$$_{-0.10}^{+0.11}$ & $0.84$$_{-0.09}^{+0.09}$ & $4.0$$_{-2.1}^{+2.3}$ &  \\
24946 & 12:37:11.11 & 62:16:38.6 & 5.193 & 25.3 & 15.0 $\pm$ 0.4 & 9.88$_{-0.34}^{+0.15}$ & $-1.47$$_{-0.09}^{+0.09}$ & $0.48$$_{-0.11}^{+0.17}$ & $49.3$$_{-25.5}^{+34.8}$ & [1,2] \\
21405 & 12:37:15.69 & 62:15:49.8 & 5.194 & 26.2 & 14.0 $\pm$ 0.3 & 9.39$_{-0.19}^{+0.17}$ & $-1.54$$_{-0.11}^{+0.10}$ & $0.35$$_{-0.11}^{+0.13}$ & $34.1$$_{-15.4}^{+27.1}$ &  \\
25788 & 12:37:16.07 & 62:16:51.7 & 5.194 & 25.3 & 17.0 $\pm$ 0.4 & 9.50$_{-0.23}^{+0.31}$ & $-1.92$$_{-0.08}^{+0.14}$ & $0.46$$_{-0.14}^{+0.09}$ & $71.6$$_{-30.7}^{+51.7}$ &  \\
24574 & 12:37:13.73 & 62:16:33.2 & 5.195 & 28.1 & 3.1 $\pm$ 0.3 & 7.96$_{-0.17}^{+0.54}$ & $-1.75$$_{-0.16}^{+0.15}$ & $0.74$$_{-0.10}^{+0.07}$ & $1.6$$_{-0.9}^{+3.0}$ &  \\
\\
\multicolumn{11}{c}{Extended Structure 7}\\
23345 & 12:36:58.43 & 62:16:15.0 & 5.192 & 26.2 & 3.5 $\pm$ 0.3 & 8.96$_{-0.26}^{+0.44}$ & $-2.11$$_{-0.07}^{+0.06}$ & $0.80$$_{-0.34}^{+0.10}$ & $13.1$$_{-5.7}^{+14.7}$ & [2,4] \\
24252 & 12:36:57.30 & 62:16:28.6 & 5.192 & 27.2 & 1.5 $\pm$ 0.2 & 8.21$_{-0.36}^{+0.23}$ & $-2.31$$_{-0.05}^{+0.04}$ & $0.97$$_{-0.03}^{+0.01}$ & $3.2$$_{-1.6}^{+2.0}$ &  \\
22502 & 12:37:05.58 & 62:16:03.9 & 5.193 & 25.5 & 8.7 $\pm$ 0.5 & 9.58$_{-0.29}^{+0.25}$ & $-1.95$$_{-0.08}^{+0.12}$ & $0.51$$_{-0.18}^{+0.17}$ & $44.0$$_{-18.3}^{+43.6}$ &  \\
24125 & 12:36:57.36 & 62:16:26.6 & 5.193 & 27.6 & 1.9 $\pm$ 0.2 & 8.76$_{-0.26}^{+0.21}$ & $-1.91$$_{-0.10}^{+0.09}$ & $0.68$$_{-0.23}^{+0.14}$ & $2.8$$_{-1.4}^{+4.5}$ &  \\
22858 & 12:37:06.63 & 62:16:08.7 & 5.194 & 27.2 & 3.5 $\pm$ 0.3 & 8.70$_{-0.33}^{+0.23}$ & $-1.50$$_{-0.15}^{+0.17}$ & $0.64$$_{-0.07}^{+0.09}$ & $5.3$$_{-2.7}^{+4.5}$ &  \\
Dusty-16116 & 12:36:56.62 & 62:17:08.0 & 5.194 & 28.2 & 2.9 $\pm$ 0.2 & 9.09$_{-0.13}^{+0.20}$ & $-1.88$$_{-0.08}^{+0.09}$ & $0.14$$_{-0.04}^{+0.06}$ & $24.1$$_{-7.8}^{+9.5}$ & [6]  \\
22295 & 12:37:05.52 & 62:16:01.3 & 5.195 & 27.4 & 3.6 $\pm$ 0.3 & 7.73$_{-0.23}^{+0.64}$ & $-2.33$$_{-0.11}^{+0.15}$ & $0.94$$_{-0.05}^{+0.04}$ & $0.8$$_{-0.2}^{+0.5}$ & [2,4] \\
22826 & 12:37:03.90 & 62:16:08.4 & 5.195 & 27.6 & 1.3 $\pm$ 0.2 & 8.61$_{-0.30}^{+0.30}$ & $-2.14$$_{-0.08}^{+0.09}$ & $0.82$$_{-0.13}^{+0.13}$ & $4.9$$_{-1.9}^{+4.3}$ &  \\
16423 & 12:37:03.28 & 62:14:53.4 & 5.200 & 27.6 & 3.0 $\pm$ 0.2 & 8.15$_{-0.32}^{+0.29}$ & $-1.83$$_{-0.14}^{+0.13}$ & $0.63$$_{-0.12}^{+0.11}$ & $2.5$$_{-1.1}^{+2.1}$ &  \\
\\
\multicolumn{11}{c}{Extended Structure 8}\\
16542 & 12:36:59.68 & 62:14:54.4 & 5.178 & 25.3 & 12.0 $\pm$ 0.5 & 10.01$_{-0.12}^{+0.11}$ & $-1.02$$_{-0.13}^{+0.11}$ & $0.66$$_{-0.14}^{+0.09}$ & $49.7$$_{-31.0}^{+54.6}$ &  \\
14125 & 12:36:50.06 & 62:14:28.7 & 5.182 & 27.5 & 1.5 $\pm$ 0.2 & 8.48$_{-0.19}^{+0.17}$ & $-2.04$$_{-0.12}^{+0.09}$ & $0.74$$_{-0.14}^{+0.12}$ & $5.1$$_{-2.2}^{+2.8}$ &  \\
20034 & 12:36:55.50 & 62:15:32.8 & 5.187 & 26.5 & 10.0 $\pm$ 0.2 & 8.27$_{-0.20}^{+0.52}$ & $-2.06$$_{-0.13}^{+0.06}$ & $0.80$$_{-0.08}^{+0.07}$ & $4.1$$_{-1.7}^{+6.8}$ & [1,2] \\
19533 & 12:36:56.35 & 62:15:27.0 & 5.188 & 25.3 & 13.0 $\pm$ 0.3 & 9.76$_{-0.31}^{+0.19}$ & $-1.49$$_{-0.11}^{+0.08}$ & $0.46$$_{-0.19}^{+0.17}$ & $85.4$$_{-51.9}^{+86.1}$ &  \\
21166 & 12:36:52.21 & 62:15:47.2 & 5.188 & 27.4 & 1.5 $\pm$ 0.2 & 8.52$_{-0.19}^{+0.18}$ & $-2.24$$_{-0.09}^{+0.09}$ & $0.87$$_{-0.26}^{+0.07}$ & $4.5$$_{-2.2}^{+3.5}$ &  \\
22601 & 12:36:52.34 & 62:16:05.1 & 5.189 & 25.3 & 8.1 $\pm$ 0.5 & 9.38$_{-0.26}^{+0.50}$ & $-1.86$$_{-0.10}^{+0.09}$ & $0.72$$_{-0.26}^{+0.14}$ & $31.4$$_{-15.7}^{+32.7}$ &  \\
20499 & 12:36:49.23 & 62:15:38.6 & 5.190 & 25.1 & 18.0 $\pm$ 0.4 & 9.19$_{-0.27}^{+0.22}$ & $-2.33$$_{-0.05}^{+0.06}$ & $0.88$$_{-0.20}^{+0.07}$ & $22.1$$_{-11.7}^{+18.8}$ & [1,2] \\
21309 & 12:36:55.39 & 62:15:48.8 & 5.191 & 26.8 & 3.6 $\pm$ 0.2 & 8.71$_{-0.46}^{+0.28}$ & $-2.17$$_{-0.08}^{+0.08}$ & $0.60$$_{-0.15}^{+0.23}$ & $9.6$$_{-5.6}^{+7.5}$ & [1,2] \\

\hline\hline
\end{tabular}
\end{table*}

\begin{table*}
\centering
\caption{Catalogs of Star-Forming Galaxies at $z\sim5.15$-5.32 that are part of the overdensity in the FRESCO GOODS North Field.}
\label{tab:sample3}
\begin{tabular}{c|c|c|c|c|c|c|c|c|c|c|c|c} \hline
     &    &     &            & $m_{210}$ & $f_{H\alpha}$\\
  ID & RA & DEC & $z_{{\rm spec}}$ & [mag] & [10$^{-18}$ ergs/s/cm$^2$] & log(M$_\star$) & $\beta$ & A$_V$ & SFR &Ref$^a$ \\\hline\hline
\multicolumn{11}{c}{Extended Structure 9}\\
26025 & 12:36:43.38 & 62:16:56.1 & 5.186 & 26.2 & 8.9 $\pm$ 0.3 & 9.39$_{-0.20}^{+0.18}$ & $-1.65$$_{-0.12}^{+0.11}$ & $0.45$$_{-0.09}^{+0.13}$ & $31.4$$_{-11.7}^{+14.8}$ &  \\
24928 & 12:36:40.75 & 62:16:38.3 & 5.188 & 25.1 & 7.8 $\pm$ 0.4 & 9.84$_{-0.12}^{+0.14}$ & $-1.20$$_{-0.10}^{+0.11}$ & $0.80$$_{-0.14}^{+0.05}$ & $38.5$$_{-15.5}^{+24.1}$ &  \\
24849 & 12:36:40.71 & 62:16:37.4 & 5.190 & 27.2 & 2.4 $\pm$ 0.2 & 8.91$_{-0.41}^{+0.27}$ & $-1.51$$_{-0.16}^{+0.12}$ & $0.66$$_{-0.19}^{+0.18}$ & $6.1$$_{-3.5}^{+4.9}$ &  \\
26502 & 12:36:47.26 & 62:17:04.2 & 5.191 & 26.0 & 2.6 $\pm$ 0.2 & 9.55$_{-0.27}^{+0.28}$ & $-2.14$$_{-0.07}^{+0.06}$ & $0.65$$_{-0.40}^{+0.22}$ & $30.5$$_{-17.9}^{+62.4}$ &  \\
28034 & 12:36:52.80 & 62:17:36.2 & 5.196 & 28.1 & 0.7 $\pm$ 0.1 & 8.07$_{-0.19}^{+0.30}$ & $-2.35$$_{-0.08}^{+0.13}$ & $0.86$$_{-0.16}^{+0.10}$ & $2.4$$_{-0.9}^{+1.7}$ &  \\
\\
\multicolumn{11}{c}{Extended Structure 10}\\
29679 & 12:36:39.72 & 62:18:24.8 & 5.192 & 25.7 & 4.4 $\pm$ 0.3 & 9.58$_{-0.28}^{+0.28}$ & $-1.99$$_{-0.09}^{+0.07}$ & $0.57$$_{-0.34}^{+0.33}$ & $39.5$$_{-26.2}^{+88.1}$ &  \\
29651 & 12:36:39.67 & 62:18:23.8 & 5.194 & 27.6 & 1.1 $\pm$ 0.1 & 8.32$_{-0.17}^{+0.23}$ & $-2.11$$_{-0.15}^{+0.14}$ & $0.76$$_{-0.17}^{+0.10}$ & $4.6$$_{-1.5}^{+2.6}$ &  \\
29544 & 12:36:39.98 & 62:18:20.2 & 5.195 & 25.5 & 10.0 $\pm$ 0.3 & 9.97$_{-0.39}^{+0.38}$ & $-1.24$$_{-0.12}^{+0.12}$ & $0.43$$_{-0.17}^{+0.24}$ & $60.1$$_{-35.6}^{+109.7}$ &  \\
\\
\multicolumn{11}{c}{Extended Structure 11}\\
6921 & 12:36:59.18 & 62:12:53.7 & 5.208 & 24.9 & 13.0 $\pm$ 0.3 & 9.89$_{-0.52}^{+0.26}$ & $-1.22$$_{-0.08}^{+0.09}$ & $0.54$$_{-0.13}^{+0.26}$ & $60.7$$_{-24.4}^{+66.4}$ &  \\
9370 & 12:37:03.31 & 62:13:31.4 & 5.212 & 25.7 & 20.0 $\pm$ 0.4 & 9.48$_{-0.22}^{+0.14}$ & $-1.94$$_{-0.07}^{+0.08}$ & $0.42$$_{-0.09}^{+0.10}$ & $57.6$$_{-23.8}^{+32.5}$ & [1,4] \\
3374 & 12:37:00.92 & 62:11:49.3 & 5.219 & 26.9 & 4.9 $\pm$ 0.2 & 8.80$_{-0.46}^{+0.30}$ & $-1.17$$_{-0.18}^{+0.20}$ & $0.53$$_{-0.12}^{+0.14}$ & $8.6$$_{-4.5}^{+8.9}$ &  \\
5204 & 12:37:08.53 & 62:12:24.5 & 5.219 & 25.5 & 5.7 $\pm$ 0.2 & 9.87$_{-0.27}^{+0.35}$ & $-1.89$$_{-0.08}^{+0.11}$ & $0.59$$_{-0.47}^{+0.21}$ & $57.5$$_{-36.8}^{+329.2}$ &  \\
3023 & 12:37:02.80 & 62:11:41.7 & 5.220 & --- & 4.3 $\pm$ 0.2 & 9.24$_{-0.33}^{+0.33}$ & $-1.79$$_{-0.13}^{+0.17}$ & $0.58$$_{-0.20}^{+0.18}$ & $18.2$$_{-12.1}^{+19.1}$ &  \\
5396 & 12:37:01.03 & 62:12:27.7 & 5.221 & 26.6 & 3.7 $\pm$ 0.2 & 9.05$_{-0.21}^{+0.22}$ & $-1.73$$_{-0.10}^{+0.10}$ & $0.56$$_{-0.23}^{+0.18}$ & $17.0$$_{-10.6}^{+13.4}$ &  \\
Dusty-4014  & 12:37:12.03 & 62:12:43.4 & 5.221 & --- & --- & $>$9.8$^c$ & ---$^c$ & ---$^c$ & ---$^c$ & [5,6] \\ 
4927 & 12:36:59.79 & 62:12:18.7 & 5.222 & 23.9 & 35.0 $\pm$ 0.7 & 10.42$_{-0.30}^{+0.19}$ & $-1.69$$_{-0.09}^{+0.16}$ & $0.62$$_{-0.15}^{+0.13}$ & $103.7$$_{-64.1}^{+190.0}$ &  \\
5702 & 12:37:03.94 & 62:12:32.9 & 5.223 & 25.2 & 8.2 $\pm$ 0.3 & 9.95$_{-0.27}^{+0.12}$ & $-2.01$$_{-0.04}^{+0.06}$ & $0.61$$_{-0.22}^{+0.14}$ & $56.7$$_{-25.0}^{+71.5}$ &  \\
5467 & 12:37:02.73 & 62:12:28.8 & 5.227 & 25.4 & 19.0 $\pm$ 0.4 & 9.68$_{-0.25}^{+0.15}$ & $-1.55$$_{-0.09}^{+0.09}$ & $0.53$$_{-0.15}^{+0.11}$ & $55.4$$_{-16.0}^{+53.2}$ & [2] \\
\\
\multicolumn{11}{c}{Extended Structure 12}\\
11160 & 12:37:18.76 & 62:13:54.1 & 5.222 & 27.9 & 1.7 $\pm$ 0.2 & 8.85$_{-0.41}^{+0.27}$ & $-1.76$$_{-0.16}^{+0.15}$ & $0.71$$_{-0.33}^{+0.15}$ & $4.8$$_{-2.7}^{+8.2}$ &  \\
14954 & 12:37:21.87 & 62:14:36.9 & 5.222 & 27.4 & 2.0 $\pm$ 0.3 & 8.40$_{-0.52}^{+0.38}$ & $-2.18$$_{-0.12}^{+0.13}$ & $0.92$$_{-0.05}^{+0.05}$ & $2.6$$_{-1.1}^{+1.4}$ &  \\
19541 & 12:37:26.47 & 62:15:27.1 & 5.222 & 26.5 & 4.6 $\pm$ 0.3 & 8.91$_{-0.24}^{+0.24}$ & $-2.10$$_{-0.09}^{+0.08}$ & $0.62$$_{-0.21}^{+0.16}$ & $14.1$$_{-5.2}^{+9.2}$ &  \\
14421 & 12:37:20.58 & 62:14:31.3 & 5.225 & 26.6 & 5.1 $\pm$ 0.3 & 8.77$_{-0.25}^{+0.25}$ & $-1.77$$_{-0.13}^{+0.18}$ & $0.60$$_{-0.10}^{+0.16}$ & $9.5$$_{-4.9}^{+6.4}$ &  \\
15403 & 12:37:21.78 & 62:14:41.4 & 5.225 & 26.9 & 4.8 $\pm$ 0.3 & 8.65$_{-0.22}^{+0.20}$ & $-2.02$$_{-0.10}^{+0.08}$ & $0.60$$_{-0.11}^{+0.15}$ & $8.6$$_{-3.3}^{+4.8}$ &  \\
\\
\multicolumn{11}{c}{Extended Structure 13}\\
26892 & 12:36:34.87 & 62:17:11.7 & 5.259 & 27.2 & 1.1 $\pm$ 0.2 & 8.63$_{-0.27}^{+0.23}$ & $-2.13$$_{-0.10}^{+0.10}$ & $0.79$$_{-0.24}^{+0.17}$ & $3.9$$_{-2.7}^{+6.0}$ &  \\
21513 & 12:36:34.97 & 62:15:51.4 & 5.267 & 26.1 & 3.8 $\pm$ 0.2 & 9.00$_{-0.24}^{+0.59}$ & $-1.85$$_{-0.08}^{+0.09}$ & $0.74$$_{-0.20}^{+0.11}$ & $12.6$$_{-5.0}^{+15.2}$ &  \\
26742 & 12:36:34.19 & 62:17:08.6 & 5.267 & 26.3 & 9.6 $\pm$ 0.3 & 8.63$_{-0.42}^{+0.25}$ & $-2.23$$_{-0.10}^{+0.11}$ & $0.86$$_{-0.08}^{+0.05}$ & $4.6$$_{-1.7}^{+3.4}$ & [2] \\
26941 & 12:36:31.44 & 62:17:12.6 & 5.269 & 26.2 & 4.2 $\pm$ 0.3 & 9.53$_{-0.91}^{+0.16}$ & $-1.09$$_{-0.11}^{+0.19}$ & $0.63$$_{-0.23}^{+0.16}$ & $12.7$$_{-4.7}^{+33.1}$ &  \\
27133 & 12:36:38.35 & 62:17:16.7 & 5.277 & 26.9 & 2.2 $\pm$ 0.3 & 8.59$_{-0.24}^{+0.18}$ & $-2.06$$_{-0.12}^{+0.10}$ & $0.85$$_{-0.10}^{+0.07}$ & $7.4$$_{-2.8}^{+3.6}$ &  \\
\\
\multicolumn{11}{c}{Extended Structure 14}\\
15414 & 12:36:32.08 & 62:14:42.0 & 5.259 & 27.5 & 1.7 $\pm$ 0.2 & 8.72$_{-0.38}^{+0.30}$ & $-1.80$$_{-0.13}^{+0.17}$ & $0.57$$_{-0.17}^{+0.19}$ & $7.3$$_{-4.1}^{+6.3}$ &  \\
15714 & 12:36:30.19 & 62:14:45.1 & 5.260 & 27.0 & 1.9 $\pm$ 0.2 & 9.13$_{-0.28}^{+0.15}$ & $-1.76$$_{-0.15}^{+0.17}$ & $0.48$$_{-0.20}^{+0.25}$ & $10.6$$_{-7.4}^{+16.7}$ &  \\
19203 & 12:36:31.94 & 62:15:23.6 & 5.268 & 27.8 & 2.7 $\pm$ 0.2 & 8.21$_{-0.27}^{+0.23}$ & $-2.22$$_{-0.09}^{+0.08}$ & $0.82$$_{-0.11}^{+0.09}$ & $2.1$$_{-1.0}^{+1.4}$ &  \\
12288 & 12:36:28.30 & 62:14:07.9 & 5.269 & 27.8 & 2.0 $\pm$ 0.1 & 8.24$_{-0.28}^{+0.14}$ & $-2.30$$_{-0.11}^{+0.11}$ & $0.83$$_{-0.22}^{+0.09}$ & $2.6$$_{-1.4}^{+2.0}$ &  \\
12805 & 12:36:30.28 & 62:14:13.8 & 5.269 & 26.7 & 2.7 $\pm$ 0.3 & 8.64$_{-0.35}^{+0.33}$ & $-2.09$$_{-0.08}^{+0.07}$ & $0.76$$_{-0.17}^{+0.10}$ & $7.2$$_{-4.1}^{+5.9}$ &  \\
15715 & 12:36:30.18 & 62:14:45.4 & 5.269 & 27.0 & 2.1 $\pm$ 0.2 & 8.59$_{-0.19}^{+0.19}$ & $-2.08$$_{-0.10}^{+0.11}$ & $0.68$$_{-0.21}^{+0.12}$ & $7.5$$_{-2.9}^{+4.1}$ &  \\
17226 & 12:36:39.34 & 62:15:01.6 & 5.269 & 27.2 & 3.0 $\pm$ 0.2 & 8.07$_{-0.30}^{+0.32}$ & $-1.99$$_{-0.10}^{+0.09}$ & $0.82$$_{-0.10}^{+0.07}$ & $1.7$$_{-0.6}^{+1.0}$ &  \\
12538 & 12:36:28.58 & 62:14:10.5 & 5.271 & 25.9 & 7.6 $\pm$ 0.3 & 9.33$_{-0.22}^{+0.21}$ & $-2.18$$_{-0.08}^{+0.08}$ & $0.43$$_{-0.15}^{+0.24}$ & $39.2$$_{-19.1}^{+28.7}$ &  \\
14448 & 12:36:29.80 & 62:14:32.0 & 5.271 & 26.8 & 3.8 $\pm$ 0.2 & 9.04$_{-0.74}^{+0.18}$ & $-2.01$$_{-0.11}^{+0.10}$ & $0.60$$_{-0.25}^{+0.23}$ & $8.7$$_{-5.3}^{+12.3}$ &  \\
13292 & 12:36:28.81 & 62:14:19.2 & 5.273 & 26.2 & 7.7 $\pm$ 0.3 & 9.12$_{-0.20}^{+0.23}$ & $-2.05$$_{-0.07}^{+0.08}$ & $0.38$$_{-0.13}^{+0.21}$ & $24.3$$_{-10.9}^{+18.6}$ &  \\
13254 & 12:36:28.71 & 62:14:18.8 & 5.275 & 25.9 & 5.3 $\pm$ 0.4 & 9.28$_{-0.20}^{+0.19}$ & $-2.08$$_{-0.06}^{+0.07}$ & $0.53$$_{-0.14}^{+0.23}$ & $29.9$$_{-12.7}^{+15.8}$ &  \\
16223 & 12:36:37.10 & 62:14:51.1 & 5.276 & 26.5 & 4.2 $\pm$ 0.3 & 9.41$_{-0.27}^{+0.24}$ & $-1.83$$_{-0.09}^{+0.10}$ & $0.53$$_{-0.21}^{+0.31}$ & $11.5$$_{-7.5}^{+14.7}$ &  \\
\\
\multicolumn{11}{c}{Extended Structure 15}\\
13539 & 12:36:22.73 & 62:14:21.7 & 5.296 & 26.4 & 7.3 $\pm$ 0.3 & 9.58$_{-0.23}^{+0.14}$ & $-1.67$$_{-0.10}^{+0.08}$ & $0.37$$_{-0.09}^{+0.22}$ & $26.6$$_{-12.5}^{+26.2}$ &  \\
13540 & 12:36:22.78 & 62:14:22.1 & 5.297 & 26.1 & 6.0 $\pm$ 0.3 & 9.45$_{-0.57}^{+0.12}$ & $-2.03$$_{-0.07}^{+0.16}$ & $0.75$$_{-0.19}^{+0.12}$ & $14.2$$_{-7.0}^{+18.0}$ &  \\
13668 & 12:36:28.66 & 62:14:23.8 & 5.298 & 25.4 & 6.3 $\pm$ 0.3 & 9.86$_{-0.35}^{+0.14}$ & $-1.54$$_{-0.11}^{+0.08}$ & $0.58$$_{-0.24}^{+0.22}$ & $63.8$$_{-47.7}^{+86.5}$ &  \\
12799 & 12:36:20.13 & 62:14:13.7 & 5.300 & 26.9 & 4.6 $\pm$ 0.2 & 8.81$_{-0.37}^{+0.23}$ & $-2.09$$_{-0.09}^{+0.10}$ & $0.68$$_{-0.11}^{+0.11}$ & $5.8$$_{-2.6}^{+6.5}$ &  \\
17951 & 12:36:31.18 & 62:15:09.0 & 5.300 & 26.2 & 4.9 $\pm$ 0.2 & 8.87$_{-0.24}^{+0.22}$ & $-2.18$$_{-0.08}^{+0.07}$ & $0.73$$_{-0.27}^{+0.13}$ & $13.3$$_{-5.9}^{+9.0}$ &  \\
12566 & 12:36:32.81 & 62:14:10.7 & 5.303 & 25.4 & 12.0 $\pm$ 0.3 & 9.69$_{-0.34}^{+0.26}$ & $-0.94$$_{-0.08}^{+0.11}$ & $0.67$$_{-0.13}^{+0.10}$ & $29.0$$_{-12.5}^{+30.5}$ &  \\
13750 & 12:36:24.20 & 62:14:24.7 & 5.305 & 26.8 & 1.6 $\pm$ 0.2 & 9.09$_{-0.31}^{+0.27}$ & $-2.00$$_{-0.06}^{+0.14}$ & $0.76$$_{-0.25}^{+0.15}$ & $7.7$$_{-4.1}^{+8.7}$ &  \\
GN10 &  12:36:33.42 & 62:14:08.6 & 5.306 & --- & --- & $>$9.7$^c$ & ---$^c$ & ---$^c$ & ---$^c$ & [3,6]\\
  \hline\hline
\end{tabular}
\end{table*}

\begin{table*}
\centering
\caption{Catalogs of Star-Forming Galaxies at $z\sim5.15$-5.32 that are part of the overdensity in the FRESCO GOODS North Field.}
\label{tab:sample4}
\begin{tabular}{c|c|c|c|c|c|c|c|c|c|c|c|c} \hline
     &    &     &            & $m_{210}$ & $f_{H\alpha}$\\
  ID & RA & DEC & $z_{{\rm spec}}$ & [mag] & [10$^{-18}$ ergs/s/cm$^2$] & log(M$_\star$) & $\beta$ & A$_V$ & SFR &Ref$^a$ \\\hline\hline
\multicolumn{11}{c}{Extended Structure 16}\\
27663 & 12:36:29.26 & 62:17:27.6 & 5.300 & 26.8 & 2.2 $\pm$ 0.2 & 9.24$_{-0.40}^{+0.30}$ & $-1.67$$_{-0.14}^{+0.13}$ & $0.59$$_{-0.26}^{+0.13}$ & $12.1$$_{-7.1}^{+28.2}$ &  \\
28064 & 12:36:29.10 & 62:17:36.9 & 5.300 & 29.6 & 2.2 $\pm$ 0.1 & 8.01$_{-0.46}^{+0.45}$ & $-1.80$$_{-0.17}^{+0.21}$ & $0.68$$_{-0.16}^{+0.12}$ & $1.1$$_{-0.7}^{+2.3}$ &  \\
28057 & 12:36:31.65 & 62:17:36.8 & 5.303 & 27.2 & 1.4 $\pm$ 0.1 & 8.42$_{-0.25}^{+0.15}$ & $-1.82$$_{-0.15}^{+0.10}$ & $0.80$$_{-0.18}^{+0.08}$ & $5.2$$_{-2.2}^{+2.3}$ &  \\
\\
\multicolumn{11}{c}{Extended Structure 17}\\
22442 & 12:37:19.75 & 62:16:02.8 & 5.297 & 26.4 & 2.6 $\pm$ 0.2 & 9.46$_{-0.42}^{+0.26}$ & $-1.89$$_{-0.06}^{+0.10}$ & $0.58$$_{-0.22}^{+0.22}$ & $22.3$$_{-10.2}^{+14.8}$ &  \\
23444 & 12:37:19.58 & 62:16:16.2 & 5.299 & 26.7 & 5.0 $\pm$ 0.2 & 9.05$_{-0.21}^{+0.19}$ & $-1.10$$_{-0.14}^{+0.14}$ & $0.54$$_{-0.15}^{+0.11}$ & $19.6$$_{-8.2}^{+11.4}$ &  \\
19151 & 12:37:18.77 & 62:15:22.7 & 5.305 & 24.5 & 17.0 $\pm$ 0.5 & 10.00$_{-0.26}^{+0.35}$ & $-1.43$$_{-0.15}^{+0.08}$ & $0.69$$_{-0.12}^{+0.08}$ & $100.5$$_{-53.4}^{+100.2}$ &  \\
16469 & 12:37:20.03 & 62:14:53.3 & 5.308 & 26.2 & 6.0 $\pm$ 0.4 & 9.59$_{-0.54}^{+0.28}$ & $-1.05$$_{-0.26}^{+0.15}$ & $0.66$$_{-0.17}^{+0.14}$ & $11.1$$_{-5.5}^{+12.4}$ &  \\
\\
\multicolumn{11}{c}{Extended Structure 18}\\
315 & 12:36:39.31 & 62:10:03.4 & 5.302 & 26.0 & 6.2 $\pm$ 0.2 & 9.10$_{-0.51}^{+0.22}$ & $-1.70$$_{-0.15}^{+0.12}$ & $0.66$$_{-0.21}^{+0.12}$ & $20.6$$_{-13.6}^{+13.6}$ &  \\
69 & 12:36:36.44 & 62:09:36.8 & 5.309 & 27.3 & 0.8 $\pm$ 0.1 & 8.64$_{-0.40}^{+0.16}$ & $-2.20$$_{-0.09}^{+0.12}$ & $0.75$$_{-0.33}^{+0.17}$ & $5.6$$_{-2.5}^{+6.3}$ &  \\
67 & 12:36:40.27 & 62:09:36.7 & 5.313 & --- & 2.6 $\pm$ 0.1 & 9.09$_{-0.61}^{+0.54}$ & $-2.01$$_{-0.08}^{+0.16}$ & $0.75$$_{-0.49}^{+0.17}$ & $8.3$$_{-4.1}^{+38.4}$ &  \\
\\
\multicolumn{11}{c}{Outside Extended Structures}\\
13227 & 12:36:58.22 & 62:14:18.7 & 5.151 & 27.7 & 1.2 $\pm$ 0.2 & 8.51$_{-0.30}^{+0.25}$ & $-1.92$$_{-0.15}^{+0.16}$ & $0.73$$_{-0.28}^{+0.16}$ & $4.7$$_{-2.5}^{+4.9}$ &  \\
2152 & 12:36:25.39 & 62:11:19.8 & 5.161 & 25.6 & 15.0 $\pm$ 0.3 & 9.57$_{-0.49}^{+0.25}$ & $-1.27$$_{-0.11}^{+0.15}$ & $0.53$$_{-0.23}^{+0.14}$ & $31.5$$_{-15.4}^{+45.1}$ &  \\
2096 & 12:36:24.97 & 62:11:18.2 & 5.163 & 25.7 & 5.3 $\pm$ 0.4 & 9.79$_{-0.27}^{+0.18}$ & $-1.52$$_{-0.20}^{+0.23}$ & $0.50$$_{-0.17}^{+0.23}$ & $37.5$$_{-25.9}^{+72.0}$ &  \\
23280 & 12:37:22.17 & 62:16:14.0 & 5.165 & 26.9 & 4.1 $\pm$ 0.3 & 8.51$_{-0.41}^{+0.31}$ & $-1.57$$_{-0.27}^{+0.20}$ & $0.60$$_{-0.14}^{+0.18}$ & $4.9$$_{-2.5}^{+3.6}$ &  \\
19527 & 12:36:38.42 & 62:15:27.2 & 5.189 & 25.7 & 5.1 $\pm$ 0.4 & 9.78$_{-0.07}^{+0.12}$ & $-1.17$$_{-0.15}^{+0.10}$ & $0.72$$_{-0.26}^{+0.09}$ & $31.0$$_{-16.2}^{+41.6}$ & [2] \\
29562 & 12:36:52.33 & 62:18:20.9 & 5.191 & 27.8 & 1.4 $\pm$ 0.2 & 8.26$_{-0.23}^{+0.31}$ & $-2.27$$_{-0.07}^{+0.10}$ & $0.91$$_{-0.12}^{+0.07}$ & $2.0$$_{-1.2}^{+2.5}$ &  \\
8937 & 12:37:17.34 & 62:13:25.2 & 5.191 & 25.8 & 7.3 $\pm$ 0.3 & 9.32$_{-0.44}^{+0.21}$ & $-2.04$$_{-0.07}^{+0.09}$ & $0.77$$_{-0.22}^{+0.11}$ & $15.1$$_{-7.2}^{+16.3}$ & [2] \\
8947 & 12:37:17.16 & 62:13:25.4 & 5.196 & 27.8 & 1.1 $\pm$ 0.1 & 8.83$_{-0.26}^{+0.35}$ & $-1.61$$_{-0.17}^{+0.18}$ & $0.55$$_{-0.29}^{+0.26}$ & $7.2$$_{-5.2}^{+12.0}$ &  \\
10911 & 12:36:24.08 & 62:13:51.2 & 5.197 & 26.3 & 3.6 $\pm$ 0.3 & 9.40$_{-0.29}^{+0.12}$ & $-1.96$$_{-0.04}^{+0.07}$ & $0.64$$_{-0.27}^{+0.09}$ & $14.1$$_{-5.6}^{+13.5}$ & [2] \\
9073 & 12:37:01.85 & 62:13:27.6 & 5.202 & 26.6 & 2.8 $\pm$ 0.4 & 9.13$_{-0.39}^{+0.38}$ & $-1.76$$_{-0.13}^{+0.12}$ & $0.65$$_{-0.09}^{+0.18}$ & $11.0$$_{-6.6}^{+9.6}$ &  \\
74 & 12:36:40.52 & 62:09:37.9 & 5.207 & --- & 2.1 $\pm$ 0.2 & 8.41$_{-0.49}^{+0.44}$ & $-2.04$$_{-0.15}^{+0.17}$ & $0.82$$_{-0.11}^{+0.09}$ & $2.5$$_{-1.2}^{+2.7}$ &  \\
23026 & 12:36:19.92 & 62:16:10.9 & 5.208 & 27.6 & 2.1 $\pm$ 0.2 & 8.61$_{-0.38}^{+0.21}$ & $-2.02$$_{-0.11}^{+0.15}$ & $0.66$$_{-0.33}^{+0.16}$ & $4.6$$_{-2.5}^{+7.6}$ &  \\
16253 & 12:37:01.00 & 62:14:51.6 & 5.209 & 28.2 & 1.1 $\pm$ 0.2 & 7.99$_{-0.49}^{+0.34}$ & $-1.74$$_{-0.18}^{+0.20}$ & $0.75$$_{-0.20}^{+0.10}$ & $1.4$$_{-0.7}^{+2.1}$ &  \\
10200 & 12:36:22.49 & 62:13:42.3 & 5.216 & 27.6 & 6.9 $\pm$ 0.2 & 7.92$_{-0.19}^{+0.22}$ & $-2.18$$_{-0.14}^{+0.13}$ & $0.75$$_{-0.07}^{+0.10}$ & $1.6$$_{-0.8}^{+0.8}$ &  \\
28662 & 12:37:06.58 & 62:17:53.2 & 5.216 & 27.4 & 1.3 $\pm$ 0.2 & 8.58$_{-0.49}^{+0.19}$ & $-2.15$$_{-0.10}^{+0.13}$ & $0.76$$_{-0.23}^{+0.15}$ & $3.8$$_{-2.1}^{+4.0}$ &  \\
6061 & 12:37:13.38 & 62:12:39.1 & 5.220 & --- & 11.0 $\pm$ 0.3 & 9.51$_{-0.26}^{+0.10}$ & $-2.14$$_{-0.09}^{+0.09}$ & $0.64$$_{-0.22}^{+0.19}$ & $57.9$$_{-27.2}^{+24.0}$ & [2] \\
2930 & 12:36:22.49 & 62:11:39.5 & 5.222 & 23.5 & 5.0 $\pm$ 0.2 & 11.07$_{-0.20}^{+0.13}$ & $-0.84$$_{-0.16}^{+0.12}$ & $0.45$$_{-0.15}^{+0.34}$ & $224.2$$_{-187.2}^{+497.3}$ &  \\
514 & 12:36:43.77 & 62:10:18.0 & 5.223 & 27.5 & 2.2 $\pm$ 0.2 & 8.55$_{-0.32}^{+0.34}$ & $-1.96$$_{-0.16}^{+0.16}$ & $0.74$$_{-0.20}^{+0.17}$ & $3.7$$_{-1.9}^{+3.2}$ &  \\
8544 & 12:37:14.58 & 62:13:18.9 & 5.223 & 26.1 & 4.8 $\pm$ 0.2 & 9.41$_{-0.21}^{+0.34}$ & $-1.44$$_{-0.11}^{+0.13}$ & $0.51$$_{-0.12}^{+0.15}$ & $43.3$$_{-22.3}^{+26.2}$ &  \\
21767 & 12:37:00.98 & 62:15:54.5 & 5.224 & 28.8 & 1.1 $\pm$ 0.1 & 7.80$_{-0.70}^{+0.60}$ & $-1.91$$_{-0.29}^{+0.21}$ & $0.74$$_{-0.27}^{+0.16}$ & $0.4$$_{-0.2}^{+1.6}$ &  \\
24801 & 12:37:00.32 & 62:16:36.7 & 5.224 & 27.6 & 1.3 $\pm$ 0.2 & 8.65$_{-0.37}^{+0.19}$ & $-2.10$$_{-0.11}^{+0.08}$ & $0.72$$_{-0.31}^{+0.19}$ & $4.7$$_{-2.4}^{+5.6}$ &  \\
11969 & 12:36:27.57 & 62:14:05.5 & 5.227 & 26.6 & 2.2 $\pm$ 0.3 & 8.07$_{-0.24}^{+0.57}$ & $-2.22$$_{-0.09}^{+0.08}$ & $0.91$$_{-0.07}^{+0.05}$ & $2.5$$_{-1.1}^{+4.0}$ &  \\
22259 & 12:37:06.39 & 62:16:00.9 & 5.227 & 27.9 & 2.0 $\pm$ 0.3 & 8.24$_{-0.33}^{+0.42}$ & $-2.14$$_{-0.12}^{+0.13}$ & $0.79$$_{-0.33}^{+0.12}$ & $2.5$$_{-1.5}^{+2.7}$ &  \\
17190 & 12:37:05.38 & 62:15:01.3 & 5.234 & 27.3 & 1.7 $\pm$ 0.2 & 8.79$_{-0.45}^{+0.25}$ & $-2.10$$_{-0.09}^{+0.09}$ & $0.84$$_{-0.47}^{+0.10}$ & $4.1$$_{-1.4}^{+6.6}$ &  \\
19141 & 12:37:06.47 & 62:15:22.8 & 5.234 & 25.6 & 4.9 $\pm$ 0.4 & 9.21$_{-0.18}^{+0.25}$ & $-2.26$$_{-0.06}^{+0.08}$ & $0.82$$_{-0.25}^{+0.13}$ & $24.6$$_{-8.7}^{+29.6}$ &  \\
21117 & 12:37:22.14 & 62:15:46.2 & 5.238 & 25.3 & 5.8 $\pm$ 0.5 & 9.84$_{-0.26}^{+0.18}$ & $-1.73$$_{-0.08}^{+0.14}$ & $0.64$$_{-0.32}^{+0.15}$ & $61.2$$_{-44.9}^{+113.4}$ &  \\
21282 & 12:37:22.63 & 62:15:48.1 & 5.240 & 26.9 & 12.0 $\pm$ 0.3 & 10.82$_{-0.31}^{+0.11}$ & $-1.49$$_{-0.04}^{+0.09}$ & $0.01$$_{0.00}^{+0.01}$ & $841.2$$_{-322.7}^{+528.9}$ &  \\
7135 & 12:36:19.66 & 62:12:56.8 & 5.246 & 27.5 & 9.9 $\pm$ 0.3 & 7.57$_{-0.11}^{+0.15}$ & $-1.99$$_{-0.13}^{+0.12}$ & $0.78$$_{-0.08}^{+0.08}$ & $0.7$$_{-0.1}^{+0.3}$ &  \\
21068 & 12:36:11.21 & 62:15:45.8 & 5.247 & 28.1 & 1.4 $\pm$ 0.1 & 8.41$_{-0.53}^{+0.31}$ & $-2.06$$_{-0.11}^{+0.12}$ & $0.73$$_{-0.14}^{+0.13}$ & $2.5$$_{-1.5}^{+2.4}$ &  \\
12053 & 12:36:22.48 & 62:14:05.3 & 5.248 & 27.4 & 2.2 $\pm$ 0.2 & 8.37$_{-0.32}^{+0.38}$ & $-1.88$$_{-0.12}^{+0.11}$ & $0.61$$_{-0.15}^{+0.18}$ & $3.9$$_{-2.0}^{+3.3}$ &  \\
13670 & 12:37:27.33 & 62:14:23.6 & 5.250 & 26.8 & 2.0 $\pm$ 0.2 & 8.37$_{-0.30}^{+0.40}$ & $-2.28$$_{-0.07}^{+0.08}$ & $0.93$$_{-0.14}^{+0.04}$ & $4.4$$_{-2.2}^{+4.5}$ &  \\
20423 & 12:37:07.73 & 62:15:37.4 & 5.272 & 25.7 & 6.5 $\pm$ 0.6 & 9.56$_{-0.44}^{+0.29}$ & $-1.22$$_{-0.11}^{+0.09}$ & $0.69$$_{-0.23}^{+0.12}$ & $38.1$$_{-22.6}^{+28.3}$ &  \\
17577 & 12:37:08.47 & 62:15:05.0 & 5.277 & 26.8 & 4.5 $\pm$ 0.2 & 8.39$_{-0.38}^{+0.34}$ & $-2.28$$_{-0.07}^{+0.08}$ & $0.89$$_{-0.04}^{+0.05}$ & $3.2$$_{-1.4}^{+2.4}$ &  \\
1253 & 12:36:33.70 & 62:10:53.7 & 5.278 & 25.7 & 13.0 $\pm$ 0.4 & 8.95$_{-0.34}^{+0.34}$ & $-2.18$$_{-0.08}^{+0.10}$ & $0.72$$_{-0.10}^{+0.14}$ & $10.8$$_{-7.6}^{+8.8}$ &  \\
983 & 12:36:46.13 & 62:10:44.8 & 5.280 & 29.0 & 1.9 $\pm$ 0.2 & 6.90$_{-0.10}^{+0.16}$ & $-2.43$$_{-0.05}^{+0.08}$ & $0.96$$_{-0.04}^{+0.03}$ & $0.1$$_{-0.0}^{+0.1}$ &  \\
12210 & 12:36:56.60 & 62:14:07.1 & 5.284 & 27.2 & 2.9 $\pm$ 0.2 & 7.65$_{-0.39}^{+1.00}$ & $-2.43$$_{-0.09}^{+0.13}$ & $0.95$$_{-0.05}^{+0.04}$ & $0.4$$_{-0.1}^{+0.3}$ &  \\
29634 & 12:36:39.58 & 62:18:23.1 & 5.288 & 27.3 & 1.8 $\pm$ 0.2 & 8.22$_{-0.22}^{+0.19}$ & $-1.94$$_{-0.10}^{+0.09}$ & $0.81$$_{-0.09}^{+0.06}$ & $3.4$$_{-1.1}^{+1.4}$ &  \\
7294 & 12:37:13.04 & 62:12:59.5 & 5.292 & 27.9 & 10.0 $\pm$ 0.2 & 7.72$_{-0.21}^{+0.26}$ & $-1.48$$_{-0.12}^{+0.12}$ & $0.60$$_{-0.13}^{+0.15}$ & $1.0$$_{-0.3}^{+0.8}$ &  \\
8583 & 12:36:54.23 & 62:13:19.6 & 5.295 & 26.6 & 2.2 $\pm$ 0.2 & 9.21$_{-0.50}^{+0.26}$ & $-2.06$$_{-0.08}^{+0.06}$ & $0.68$$_{-0.43}^{+0.15}$ & $12.1$$_{-5.0}^{+24.2}$ &  \\
  \hline\hline
\end{tabular}
\end{table*}

\begin{table*}
\centering
\caption{Catalogs of Star-Forming Galaxies at $z\sim5.15$-5.32 that are part of the overdensity in the FRESCO GOODS North Field.}
\label{tab:sample5}
\begin{tabular}{c|c|c|c|c|c|c|c|c|c|c|c|c} \hline
     &    &     &            & $m_{210}$ & $f_{H\alpha}$\\
  ID & RA & DEC & $z_{{\rm spec}}$ & [mag] & [10$^{-18}$ ergs/s/cm$^2$] & log(M$_\star$) & $\beta$ & A$_V$ & SFR &Ref$^a$ \\\hline\hline
  \multicolumn{11}{c}{Outside Extended Structures}\\
  11214 & 12:36:35.85 & 62:13:54.7 & 5.296 & 26.0 & 7.2 $\pm$ 0.2 & 9.41$_{-0.25}^{+0.21}$ & $-1.78$$_{-0.09}^{+0.09}$ & $0.63$$_{-0.11}^{+0.14}$ & $14.7$$_{-4.7}^{+11.7}$ &  \\
2243 & 12:36:52.12 & 62:11:22.7 & 5.296 & 28.0 & 2.4 $\pm$ 0.3 & 8.18$_{-0.30}^{+0.43}$ & $-2.29$$_{-0.15}^{+0.10}$ & $0.93$$_{-0.13}^{+0.05}$ & $2.0$$_{-0.6}^{+1.6}$ &  \\
10647 & 12:36:12.62 & 62:13:47.6 & 5.297 & 26.4 & 2.4 $\pm$ 0.1 & 9.34$_{-0.30}^{+0.29}$ & $-1.78$$_{-0.12}^{+0.11}$ & $0.49$$_{-0.20}^{+0.31}$ & $23.4$$_{-13.0}^{+25.8}$ &  \\
6255 & 12:36:54.32 & 62:12:42.9 & 5.307 & 25.7 & 6.0 $\pm$ 0.4 & 9.77$_{-0.16}^{+0.16}$ & $-1.49$$_{-0.07}^{+0.11}$ & $0.61$$_{-0.23}^{+0.18}$ & $31.5$$_{-17.1}^{+42.2}$ &  \\
17397 & 12:37:24.83 & 62:15:02.7 & 5.310 & 24.7 & 42.0 $\pm$ 0.6 & 9.82$_{-0.18}^{+0.20}$ & $-1.35$$_{-0.11}^{+0.10}$ & $0.44$$_{-0.09}^{+0.11}$ & $113.8$$_{-35.8}^{+60.3}$ &  \\
15255 & 12:36:03.80 & 62:14:39.7 & 5.315 & 26.0 & 3.8 $\pm$ 0.3 & 9.15$_{-0.28}^{+0.37}$ & $-2.19$$_{-0.10}^{+0.07}$ & $0.87$$_{-0.21}^{+0.10}$ & $13.2$$_{-6.7}^{+27.5}$ &  \\
24012 & 12:36:29.52 & 62:16:25.0 & 5.316 & 27.6 & 1.5 $\pm$ 0.2 & 8.33$_{-0.53}^{+0.32}$ & $-1.92$$_{-0.14}^{+0.14}$ & $0.76$$_{-0.19}^{+0.12}$ & $1.7$$_{-0.8}^{+1.9}$ &  \\
  \hline\hline
\end{tabular}
\end{table*}

\end{document}